\documentclass[12pt]{article}
\usepackage{hfstyle}
\usepackage{amssymb,setspace,epic,eepic,graphicx}

\newcommand{\tenrm}{}
\begin{document}
\title{Performance of data networks with random links}

\author{Henryk Fuk\'s\footnote{From July 2000: Department of Mathematics, Brock University,
St. Catharines, Ontario  L2S 3A1, Canada,
\texttt{hfuks@brocku.ca}}
        \,   and Anna T. Lawniczak
      \oneaddress{
         Department of Mathematics and
         Statistics, \\University of Guelph, \\
         Guelph, Ontario N1G 2W1, Canada\\
         {and}\\
         The Fields Institute for Research\\
         in Mathematical Sciences\\
         Toronto, Ontario M5T 3J1, Canada\\
         \email{$\{$hfuks,alawnicz$\}$@fields.utoronto.ca}
       }
   }

 \date{}  

\Abstract{We investigate simplified models of computer data networks
and examine how the introduction of additional random links influences
the performance of  these networks. In general, the impact of additional
random links on the performance of the network strongly depends on the
routing algorithm used in the network. Significant performance gains can
be achieved if the routing is based on ``geometrical distance'' or
shortest path  reduced table routing. With shortest path full table
routing degradation of performance is observed. }
\maketitle
\section{Introduction}
Models of computer data networks have attracted much attention in recent
years. Generally, these models assume either regular topology of the
network, in the form of a square lattice \cite{Campos95,Deane96,Ohira98} or
a binary Cayley tree \cite{Tretyakov98}, or random graph topology \cite
{Kadirire94}. On the other hand, it has been recently demonstrated that many
technological, biological, and social networks are neither completely
regular nor completely random, being somewhere between these two extremes
\cite{Duncan98}. Regular lattices rewired to introduce small amount of
random connections, termed ``small world networks'', offer many advantages
over purely regular or purely random topologies. In particular, models of
dynamical systems based on ``small world'' lattices can often exhibit
enhanced signal propagation capabilities, as observed in epidemic models or
multi-player prisoners dilemma games played on such lattices~\cite{Duncan98}.

The purpose of this work is to investigate if a similar effect can be
achieved in simple data network models, although our approach is
different than the approach taken in~\cite{Duncan98}. We do not rewire
the network, but rather examine how introduction of additional random
links influences its performance. This is motivated by the question
whether introduction of additional links can help to decongest an
existing network.

\section{Network Models Definitions}

The purpose of the network is to transmit messages from points of origin
to destination points. In our model, we will assume that the entire
message is contained in a single ``capsule'' of information, which, by
analogy to packet-switching networks, will be simply called a \emph{packet}.
In a real packet-switching network, a single packet carries the information
``payload'', and some additional information related to the internal
structure of the network. We will ignore the information ``payload''
entirely, and assume that the packet carries only two pieces of information:
time of its creation and the destination address.

Our simulated network consists of a number of interconnected nodes. Each
node can perform two functions: of a \emph{hosts}, meaning that it can
generate and receive messages, and of a \emph{router} (message processor),
meaning that it can store and forward messages. Packets are created and
moved according to a discrete time parallel algorithm. The structure of the
considered networks and the update algorithm will be described in
subsections which follow.

\subsection{Connection Topologies}

We will consider two types of network connection topologies: a
two-dimensional square lattice $\mathcal{L=L(}L)$ and a two-dimensional
square lattice $\mathcal{L}$ with additional links added randomly, denoted
by $\mathcal{L}_{l}=\mathcal{L}_{l}(L)$. The value of the subscript $l$
gives a number of an extra links in a network and the value of $L$ gives a
number of nodes in the horizontal and vertical direction of the lattice $%
\mathcal{L}$. The lattices $\mathcal{L}$ and $\mathcal{L}_{l}$ with periodic
boundary conditions will be denoted by $\mathcal{L}^{p}$ and $\mathcal{L}%
_{l}^{p}$, respectively, and with non-periodic boundary\ conditions by $%
\mathcal{L}^{np}$ and $\mathcal{L}_{l}^{np}$, respectively. Hence, with this
notation, $\mathcal{L}_{0}^{p}=\mathcal{L}^{p}$ and $\mathcal{L}_{0}^{np}=%
\mathcal{L}^{np}$. Most of our simulations will be performed on lattices
with periodic boundary conditions.

The network \emph{hosts} and \emph{routers} are located at nodes of the
lattice $\mathcal{L}$. The position of each node on a lattice $\mathcal{L}$
is described by a discrete space variable $\mathbf{r}$, such that
\begin{equation}
\mathbf{r}=i\mathbf{c}_{x}+j\mathbf{c}_{y},
\end{equation}
where $\mathbf{c}_{x},\mathbf{c}_{y}$ are Cartesian unit vectors, and $%
i,j=1,\ldots ,L$. For each node $\mathbf{r}$ we denoted by $C(\mathbf{r})$
the set of all nodes directly connected with the node $\mathbf{r}$. Hence,
for each $\mathbf{r}\in $ $\mathcal{L}^{p}$, the set $C(\mathbf{r})$ is of
the form
\begin{equation}
C(\mathbf{r})=\{\mathbf{r}-\mathbf{c}_{x},\mathbf{r}+\mathbf{c}_{x},\mathbf{r%
}-\mathbf{c}_{y},\mathbf{r}+\mathbf{c}_{y}\}.  \label{nbh}
\end{equation}
In this case, the node $\mathbf{r}$ is connected with its four nearest
neighbours. However, for lattices with non-periodic boundary conditions $%
\mathcal{L}^{np}$ or square lattices with additional links added randomly $%
\mathcal{L}_{l}^{p}$ and $\mathcal{L}_{l}^{np}$ the form of the set $C(%
\mathbf{r})$ can be different, for some nodes $\mathbf{r,}$ from the one in (%
\ref{nbh}). For example, for nodes $\mathbf{r}$ on the boundary of a lattice
$\mathcal{L}^{np}$ the set $C(\mathbf{r})$ can contain two or three elements
only, depending on where a node $\mathbf{r}$ is located on the boundary. In
the case of a lattice $\mathcal{L}_{l}^{p}$ or $\mathcal{L}_{l}^{np}$ the
set $C(\mathbf{r})$ can contain many non-nearest neighbours nodes depending
on a number of additional links which originate from the node $\mathbf{r}$.

The extra links are constructed using the following procedure. We first
select randomly a node ${\mathbf{r}}_{1}$ on a square lattice $\mathcal{L}%
^{p}$ or $\mathcal{L}^{np}$. Next, we select randomly another node ${\mathbf{%
r}}_{2}$, different from the node ${\mathbf{r}}_{1}$, and connect these two
nodes with direct communication link. By repeating this procedure
independently $l$ times we obtain a lattice $\mathcal{L}_{l}^{p}$ or $%
\mathcal{L}_{l}^{np}$, respectively, with additional $l$ random links. It
can happen that the nodes $\mathbf{r}_{1}$ and $\mathbf{r}_{2}$ can be
selected again to form a new link. Hence, in the network there can be
several links connecting directly the same nodes. We want to emphasize that
all the connections in our models are \emph{static}, during the simulation
period they do not change. Additional random links are added before the
simulation starts, and remain unchanged.

In the networks considered here, each node maintains a queue of unlimited
length where the arriving packets are stored. The number of packets in the
queue at a node $\mathbf{r}$ at time step $k$ will be denoted by $n(\mathbf{r%
},k)$, while the total number of packets in the system at time step $k$ will
be denoted by $N(k)$,
\begin{equation}
N(k)=\sum_{{\mathbf{r}}\in \mathcal{L}}{n({\mathbf{r}},k)}.
\end{equation}
Packets stored in queues, at individual lattice nodes, must be delivered to
their destination addresses. To assess how far a given packet is from its
destination, we introduce the concept of distance between nodes. Depending
on a network connection topology we will use three metric functions to
compute the distance between nodes ${\mathbf{r}}_{1}=(i_{1},j_{1})$ and ${%
\mathbf{r}}_{2}=(i_{2},j_{2})$. Namely, we will use

\begin{enumerate}
\item  for lattices with non-periodic boundary conditions
``Manhattan'' metric
\begin{equation}
d_{M}({\mathbf{r}}_{1},{\mathbf{r}}_{2})=|i_{2}-i_{1}|+|j_{2}-j_{1}|,
\end{equation}

\item  for lattices with periodic boundary conditions periodic
``Manhattan'' metric
\begin{equation}
d_{PM}({\mathbf{r}}_{1},{\mathbf{r}}_{2})=L-\left| |i_{2}-i_{1}|-\frac{L}{2}%
\right| -\left| |j_{2}-j_{1}|-\frac{L}{2}\right| ,
\end{equation}

\item  and regardless of boundary conditions the ``shortest path'' metric $%
d_{SP}({\mathbf{r}}_{1},{\mathbf{r}}_{2})$ defined as the number of links in
the shortest path joining ${\mathbf{r}}_{1}$ and ${\mathbf{r}}_{2}$. By
``shortest'' we mean the path with the smallest number of links.
\end{enumerate}

Metric $d_{PM}$ will be used on lattices $\mathcal{L}^{p}$,
while $d_{M}$ will be used on lattices $\mathcal{L}^{np}.$ Note that on a
square lattice with no extra links $d_{SP}(%
{\mathbf{r}}_{1},{\mathbf{r}}_{2})=d_{M}({\mathbf{r}}_{1},{\mathbf{r}}_{2})$%
, and $d_{SP}({\mathbf{r}}_{1},{\mathbf{r}}_{2})=d_{PM}({\mathbf{r}}_{1},{%
\mathbf{r}}_{2})$ for lattices with non-periodic and periodic boundary
conditions, respectively. Furthermore, for each $\mathbf{r\in \mathcal{L}}%
_{l}^{\kappa }$, where $\kappa =p$ or $np$ and $l\in \{0,1,...\}$

\begin{equation}
C(\mathbf{r)=}\left\{ \mathbf{x\in \mathcal{L}}_{l}^{\kappa }
\mathbf{:} d_{SP}(\mathbf{x},\mathbf{r})=1\right\}.
\end{equation}
When irregularities such as extra links are present, $d_{SP}({\mathbf{r}}%
_{1},{\mathbf{r}}_{2})$ can be computed using one of the well known
algorithms. In our simulations, we used \emph{shortest path backward tree}
algorithm \cite{Saadawi94}.

\subsection{Routing Algorithms}

\begin{figure}[p]
\begin{center}
\includegraphics[scale=0.8]{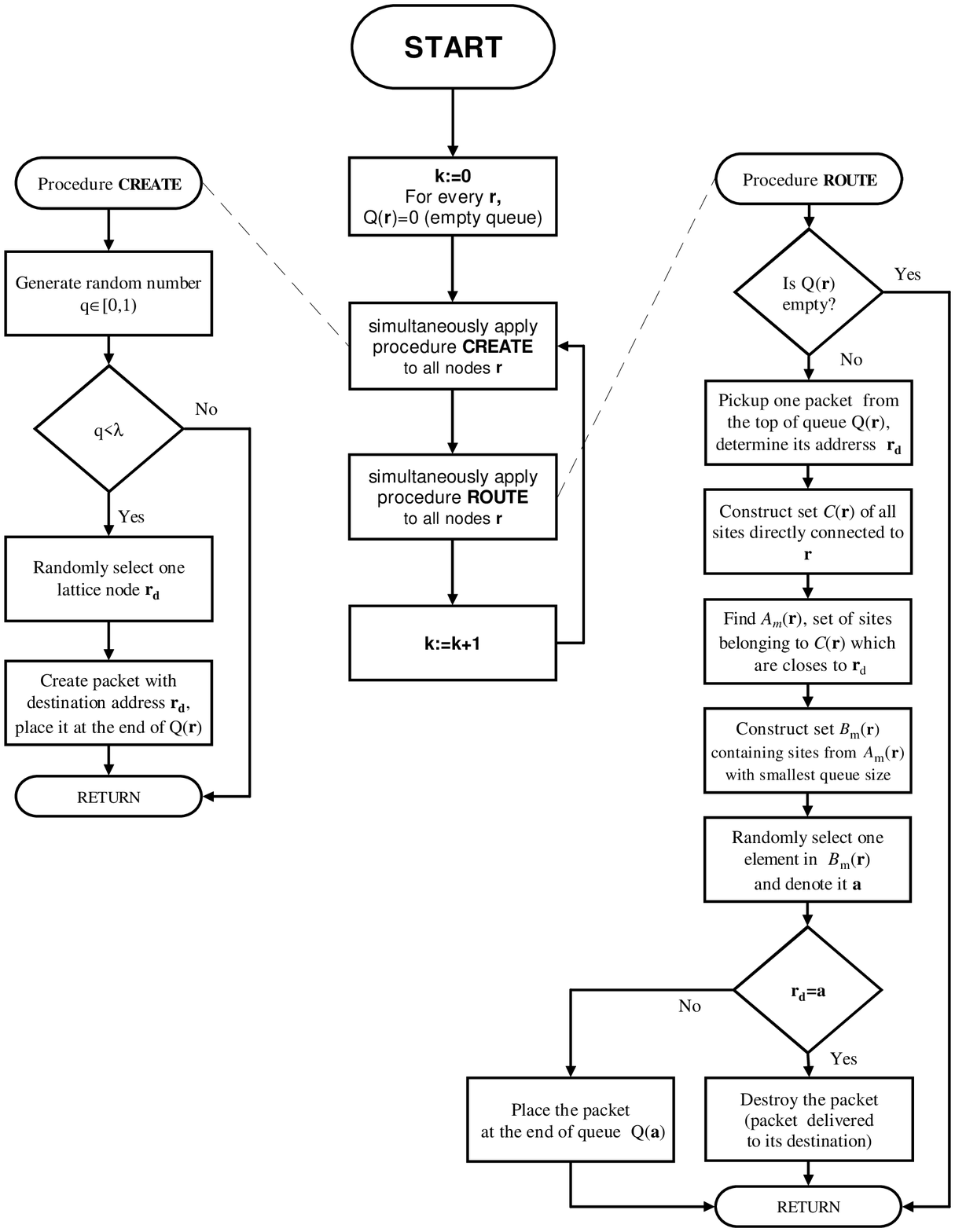}
\end{center}
\caption{Network update algorithm. Symbol $Q(\mathbf{r})$ denotes
the queue at node $\mathbf{r}$.}
\label{fig90}
\end{figure}

The dynamics of the networks are governed by the parallel update
algorithms shown Figure \ref{fig90}, similar to the algorithm used
in~\cite{Ohira98}. We start with an empty queue at each node, and with
discrete time clock $k$ set to zero. Then, the following actions are
performed in sequel:

\begin{enumerate}
\item  At each node,\ independently of the others, a packet is created with
probability $\lambda $. Its destination address is randomly selected among
all other nodes in the network with uniform probability distribution. The
newly created packet is placed at the end of the queue.

\item  At each node, one packet (or none, if the local queue is empty) is
picked up from the top of the queue and forwarded to one of its neighboring
sites according to a one of the routing algorithms to be described below.
Upon arrival, the packet is placed at the end of the appropriate queue. If
several packets arrive to a given node at the same time, then they are
placed at the end of the queue in a random order. When a packet arrives to
its destination node, it is immediately destroyed.

\item  $k$ is incremented by 1.
\end{enumerate}

This sequence of events, which constitutes a \emph{single time step update},
is then repeated arbitrary number of times. The state of the network is
observed after sub-step 3 (clock increase), but before sub-step 1 (creation
of new packets). In order to explain the routing algorithms mentioned in
sub-step 2, we will first describe one of its simplified versions.

Let us assume that we measure distance using some metric $d$, where $d$
could be any of the previously defined metrics $d_{M}$, $d_{PM}$, or $d_{SP}$%
. To decide where to forward a packet located at a node $\mathbf{r}$ with
the destination address $\mathbf{r}_{d}$, two steps are performed:

\begin{enumerate}
\item  From sites directly connected to $\mathbf{r}$, we select sites which
are closest to the destination $\mathbf{r}_{d}$ of the packet. More
formally, we construct a set $A_{\infty }(\mathbf{r})$ such that
\begin{equation}
A_{\infty }(\mathbf{r})=\{\mathbf{a}\in C(\mathbf{r}):d(\mathbf{a},\mathbf{r}%
_{d})=\min_{\mathbf{x}\in C(\mathbf{r})}d(\mathbf{x},\mathbf{r}_{d})\}
\label{seta}
\end{equation}

\item  From $A_{\infty }(\mathbf{r})$, we select a site which has the
smallest queue size. If there are several such sites, then we select one of
them randomly with uniform probability distribution. The packet is forwarded
to this site. Using a formal notation again, we could say that the packet is
forwarded to a site selected randomly and uniformly from elements of a set $%
B_{\infty }(\mathbf{r})$ defined as
\begin{equation}
B_{\infty }(\mathbf{r})=\{\mathbf{a}\in A_{\infty }(\mathbf{r}):n(\mathbf{a}%
,k)=\min_{\mathbf{x}\in A_{\infty }(\mathbf{r})}n(\mathbf{x},k)\}.
\end{equation}
\end{enumerate}

To summarize, the routing algorithm $\mathbf{R}_{\infty }$ described above
sends the packet to a site which is closest to the destination (in the sense
of the metric $d$), and if there are several such sites, then it selects
from them the one with the smallest queue. If there is still more than one
such node, random selection takes place. It is clear that each packet routed
according to the algorithm $\mathbf{R}_{\infty }$ will travel to its
destination taking the shortest possible path (shortest in the sense of the
metric $d$, not necessarily in terms of the number of time steps required to
reach the destination). In real networks, this does not always happen. In
order to allow packets to take alternative routes, not necessarily shortest
path routes, we will introduce a small modification to the routing algorithm
$\mathbf{R}_{\infty }$ described above.

The modified algorithm $\mathbf{R}_{m}$, for each node $\mathbf{r,}$ will
use instead of the set $A_{\infty }(\mathbf{r})$ a set $A_{m}(\mathbf{r})$
defined as follows. In the construction of the set $A_{m}(\mathbf{r})$
instead of minimizing distance to the destination $d(\mathbf{x},\mathbf{r}%
_{d})$, as it was done in $(\ref{seta})$, we will minimize $\Theta _{m}(d(%
\mathbf{x},\mathbf{r}_{d}))$, where
\begin{equation}
\Theta _{m}(y)=\left\{
\begin{array}{ll}
y, & \mbox{if $y < m$}, \\
m, & \mbox{otherwise},
\end{array}
\right.
\end{equation}
for a given integer $m$. Thus, the definition of the set $A_{m}(\mathbf{r})$
is
\begin{equation}
A_{m}(\mathbf{r})=\{\mathbf{a}\in C(\mathbf{r}):\Theta _{m}(d(\mathbf{a},%
\mathbf{r}_{d}))=\min_{\mathbf{x}\in C(\mathbf{r})}\Theta _{m}(d(\mathbf{x},%
\mathbf{r}_{d}))\}
\end{equation}
The above modification is equivalent to saying that nodes which are further
than $m$ distance units from the destination are treated by the routing
algorithm \textit{as if they were exactly $m$ units away from the destination%
}. If a packet is at a node $\mathbf{r}$ such that all nodes directly linked
with $\mathbf{r}$ are further than $m$ units from its destination, then the
packet will be forwarded to a site selected randomly and uniformly from the
subset of $C(\mathbf{r})$ containing the nodes with the smallest queue size
in the set $C(\mathbf{r}).$ It can happen that the selected site can be
further away from the destination than the node $\mathbf{r}$.

Therefore, introduction of the\emph{\ cutoff parameter} $m$ adds more
randomness to the network dynamics. One could also say that the destination
attracts packets, but this attractive interaction has a finite range $m$:
packets further away than $m$ units from the destination are not being
attracted.

It is also possible to relate various values of the cutoff parameter $m$ to
different types of routing schemes used in real packet-switching networks.
Assume that each node $\mathbf{r}$ maintains a table containing all possible
values of $d(\mathbf{x},\mathbf{r}_{d})$, for all possible destinations $%
\mathbf{r}_{d}$ and all nodes $\mathbf{x}\in C(\mathbf{r})$, and that
packets are routed according to this table by selecting nodes minimizing
distance, measured in the metric $d$, travelled by a packet from its origin
to its destination. Such a routing scheme is called \emph{table-driven
routing} \cite{Saadawi94} and it is equivalent to the routing algorithm $%
\mathbf{R}_{\infty }$. In this case, construction of the set $A_{\infty }(%
\mathbf{r})$ would require looking up appropriate entries in the stored
table.

Let us now define $D_{max}$ to be the largest possible distance between two
nodes in the network. When $m<D_{max}$, then for a given ${\mathbf{x}}$, we
need to store values of $d({\mathbf{x}},{\mathbf{r}}_{d})$ only for nodes ${%
\mathbf{r}_{d}}$ which are less than $m$ units of distance away -- for all
other nodes distance does not matter, since it will be treated as $m$ by the
routing algorithm. Hence, at each node $\mathbf{r}$ the routing table to be
stored is smaller than in the case when $m=D_{max}$. The routing scheme
based on this smaller routing table is called the \emph{reduced table
routing algorithm }\cite{Saadawi94} and it is equivalent to the routing
algorithm $\mathbf{R}_{m}$. In the case when $m=D_{max}$ the routing
algorithm $R_{m}=R_{\infty }.$

Finally, when $m=1$, the distances between hosts and destinations are not
considered in the routing process of packets. Therefore, there is no need to
store any table of possible paths at nodes of the network. This case
corresponds to the \emph{table-free routing algorithm }\cite{Saadawi94} in
which packets are routed randomly. Hence, this algorithm can send packets on
circuitous and long routes to their destinations. The analysis of this
routing algorithm has been done in \cite{FLW99}, where some analytical
results are also presented. At present such results are not available for
routing algorithms with $m>1$.

\section{Performance of networks with square lattice connection topology $(%
\mathcal{L}_{0}^{p},d_{PM}),$ $(\mathcal{L}_{0}^{np},d_{M})$}

In order to asses the performance of a network, graphs of \emph{delay} as a
function of \emph{presented load} are frequently used in network performance
literature \cite{Stallings98}. In our case, $\emph{delay}$ $\tau $ will be
defined as the number of time steps elapsed from the creation of a packet to
its delivery to the destination address. We will also use \emph{average delay%
} $\overline{\tau }(k)$, where the average is taken over all packets
delivered to their destination from the beginning of the simulation ($k=0$)
up to time $k$. Probability of a packet creation $\lambda $ will be used as
a measure of a \emph{presented load}.

\subsection{Full table routing}

We assume that the network topology is a square lattice $\mathcal{L}%
_{0}^{p}(L)$ with $d_{PM}$ metric and the network routing algorithm is the
full table routing algorithm $\mathbf{R}_{m}$, with $m=D_{max}$, i.e. $%
R_{m}=R_{\infty }.$ Figure \ref{fig100}a shows graphs of the average delay $%
\overline{\tau }(k)$ versus presented load $\lambda $, as measured during
simulation performed on a lattice $\mathcal{L}_{0}^{p}(50).$ The three
curves shown there correspond to different times. It is clear that beyond a
certain critical value of $\lambda =\lambda _{c}$, the average delay
drastically increases. Moreover, the average delay grows with time, which
suggests that for $\lambda >\lambda _{c}$ there is no equilibrium state. In
fact, when $\lambda >\lambda _{c}$, a typical queue size and consequently,
the number of packets in the system $N(k),$ grows without bounds, as shown
in Figure \ref{fig100}b and Figure \ref{fig900}.

It is possible to find an approximate value of the critical load $\lambda_c$
by the following argument. For $\lambda<\lambda_c$, the system reaches
steady state, and in the steady state the number of packets created per unit
time (given by $L^2\lambda$) must be equal to the number of packets
delivered per unit time. Since the average time spent in the system by a
packet is $\overline{\tau}(k)$, we can reasonably assume that $N(k)/
\overline{\tau}(k)$ packets are delivered to their destinations per unit
time, hence
\begin{equation}
\frac{N(k)}{\overline{\tau}(k)}=L^2 \lambda.  \label{little}
\end{equation}
This relationship, known as \emph{Little's law} in queuing theory \cite
{Nelson95}, holds only below the critical point, as shown in Figure~\ref
{fig300}.

For the routing algorithm $\mathbf{R}_{m}$, with $m=D_{max}$, when the
number of packets in the network is small, an individual packet is always
routed in such a way that it follows the shortest path to its destination
avoiding all occupied nodes. This means that for small $N(k)$, the average
packet delay is approximately equal to average distance from the packet's
origin to its destination, which will be called \emph{``free packet'' delay}
$\overline{\tau }_{0}$
\begin{equation}
\overline{\tau }_{0}=\frac{1}{L^{4}}\sum_{{\mathbf{r}}_{1},{\mathbf{r}}%
_{2}}d_{PM}({\mathbf{r}}_{1},{\mathbf{r}}_{2})
\end{equation}
After some algebra, this leads to
\begin{equation}
\overline{\tau}_{0}=\frac{1}{L^{4}}\sum_{i_{1},i_{2},j_{1},j_{2}=0}^{L-1}%
\left\{ L-\left| |i_{2}-i_{1}|-\frac{L}{2}\right| -\left| |j_{2}-j_{1}|-%
\frac{L}{2}\right| \right\} =\frac{L}{2}.
\end{equation}
Obviously, when the load increases, at some point the number of packets in
the network will be so large that it would not be possible to find a route
to a destination completely avoiding other packets. Assuming that packets
are approximately uniformly distributed over the entire lattice, this will
happen when all sites are occupied, i.e. when $N(k)=L^{2}$. Using (\ref
{little}) this gives an estimate of $\lambda _{c}$:
\begin{equation}
\lambda _{c}=\frac{1}{\overline{\tau }_{0}}
\end{equation}

For $\mathcal{L}_{0}^{p}(50)$ we obtain $\overline{\tau }_{0}=25$ and $%
\lambda _{c}=0.04$, in good agreement with the value obtained from
simulations $\lambda _{c}=0.039\pm 0.001$.

Quite similar calculations can be performed for a square lattice with
non-periodic boundary $\mathcal{L}_{0}^{np}(L)$. In this case,
\begin{equation}
\overline{\tau }_{0}=\frac{1}{L^{4}}\sum_{{\mathbf{r}}_{1},{\mathbf{r}}%
_{2}}d_{M}({\mathbf{r}}_{1},{\mathbf{r}}_{2})=\frac{2}{3}\frac{L^{2}-1}{L}%
\approx \frac{2}{3}L,
\end{equation}
yielding $\lambda _{c}=0.03$ for $\mathcal{L}_{0}^{np}(50).$ The measured
value of $\lambda _{c}$ for $\mathcal{L}_{0}^{np}(50)$ is $0.020\pm 0.001$,
i.e., much lower. The discrepancy is mainly due to the fact that for the
lattices $\mathcal{L}_{0}^{np}(L)$ packets are not uniformly distributed on
the lattice, having a tendency to cluster at the center. Consequently,
jamming occurs earlier than one would expect assuming uniform distribution
of packets.

\subsection{Partial table routing}

Decrease in value of the cutoff parameter $m$ has a profound effect on the
critical load. Smaller $m$ means that packets which are located further than
$m$ links from their destination move with a high degree of randomness, and
as a result, their average delay is larger. This increase of the delay can
be also seen in a plot of a single packet delay as a function of $m$ (Figure~%
\ref{fig500}). While values of $m$ close to $D_{max}$ do not significantly
change $\overline{\tau }_{0}$, values of $m$ close to $1$ result in an
increase of $\overline{\tau }_{0}$ by up to two orders of magnitude.

\section{Performance of networks based on square lattices $(\mathcal{L}%
_{l}^{p},d_{PM}),$ $(\mathcal{L}_{l}^{np},d_{M})$ with additional $l$ random
links}

\subsection{Full table routing}

Let us now consider the network dynamics governed by the routing algorithm $%
R_{\infty }$ taking place on lattices $\mathcal{L}_{l}^{p}$ and $\mathcal{L}%
_{l}^{np}$which in addition to normal nearest neighbor connections, feature $%
l$ additional links, where $l>0$. Figure~\ref{fig400} shows how addition of
random links changes the graph of delay vs. load for both non-periodic and
periodic case. We are still using $d_{M}$ and $d_{PM}$ metric for $\mathcal{L%
}_{l}^{np}$ and $\mathcal{L}_{l}^{p}$ lattice, respectively, which means
that the distance between two points ${\mathbf{r}}_{1}$ and ${\mathbf{r}}_{2}
$ is still computed using $d_{M}({\mathbf{r}}_{1},{\mathbf{r}}_{2})$ or $%
d_{PM}({\mathbf{r}}_{1},{\mathbf{r}}_{2})$ metric, respectively, even if ${%
\mathbf{r}}_{1}$ and ${\mathbf{r}}_{2}$ are directly connected by some extra
link.

As expected, addition of extra links improves performance of the network,
shifting the critical point $\lambda _{c}$ to the right (see Figures~\ref
{fig400} and \ref{fig550}). This means that the network can carry more load
without experiencing congestion. Performance improvement is more pronounced
for lattices with non-periodic boundaries, as shown in Figure~\ref{fig550}.
For example, by adding 100 random links to $50\times 50$ lattice, which
increases total number of links by $2\%$, we increase the critical load by
over $25\%$. Increasing the number of links by $8\%$ doubles the critical
load. This can be attributed to the fact that some packets can bypass
congested central area by using ``shortcuts'', and their delay decreases not
only because they have shorter distance to travel, but also because they
have avoided congestion. In the case of lattices with periodic boundaries,
packets are more uniformly spread even in the presence of additional random
links. Thus, the performance improvement is only caused by the decrease in
the distance traveled, but not by bypassing congestion, since congestions
are also uniformly spread in the case of lattices with periodic boundaries.

In the remainder of this article, we will focus our discussion on lattices
with periodic boundaries only.

\subsection{Partial table routing}

When a routing algorithm $\mathbf{R}_{m},$ with $m<D_{max},$ is used,
additional random links can significantly increase critical load, and the
relative performance gain is much larger than in the case of the full table
routing algorithm. Figure~\ref{fig600}a,b shows the relative change of the
critical load, defined by
\begin{equation}
\frac{\Delta \lambda _{c}}{\lambda _{c}}=\frac{\lambda _{c}(m,l)-\lambda
_{c}(m,0)}{\lambda _{c}(m,0)},
\end{equation}
where $\lambda _{c}(m,l)$ denotes the critical load at a given $m$ and $l$,
for two different values of $m$, $m=50$ and $m=20$. One can immediately
notice that the impact of additional links on performance of the network is
much stronger in the case of partial table routing ($m=20$) than in the case
of full table routing ($m=50)$. For example, about $50$ extra links are
sufficient to double the critical load corresponding to $m=20$, while the
same number of links has almost negligible impact on the critical load when $%
m=50$.

\section{Performance of networks based on square lattice $(\mathcal{L}%
_{l}^{p},d_{SP})$ with additional $l$ random links and $d_{SP}$ metric}

As stated before, for a square lattice without additional links, metric $%
d_{SP}$ is identical to $d_{M}$ or $d_{PM}$ metric. This is no longer true
for a square lattice with additional random links. Routing based on $d_{SP}$
metric fully utilizes shortcuts provided by additional links, significantly
decreasing ``free packet'' delay $\overline{\tau }_{0}$. One would expect
that a decrease in ``free packet'' delay will decrease also average delay,
as it was in the case of networks with $d_{M}$ and $d_{PM}$ metric. In
reality, we observe just opposite effect (Figure \ref{fig600}).

\subsection{Full table routing}

Figure~\ref{fig600}c shows how the critical load $\lambda _{c}(m,l)$ changes
when additional random links are introduced. This is shown for the network
dynamics governed by $R_{\infty }$, i.e. the full table routing algorithm
with $m=D_{max}$, on a square lattice with periodic boundaries  One can
clearly see that if the number of additional random links $l$ is below some
critical value $l_{c}(m)$ the critical load $\lambda _{c}(D_{max},l)$ is
actually \emph{smaller }than $\lambda _{c}(D_{max},0)$, in spite of
increased connectivity between nodes of the network. The performance of the
network is at its worst when just a few additional random links are added.
However, it improves with the increase of a number $l$ of additional random
links and at some critical value $l_{c}(m)$ it becomes the same as of the
network without any additional random link. When the number $l$ of
additional random links is greater than $l_{c}(m)$ an improvement in the
network performance is observed. For the network $\mathcal{L}_{500}^{p}(50)$
the critical load $\lambda _{c}(50,500)$ is almost equal to the critical
load $\lambda _{c}(50,0)$ of the network $\mathcal{L}_{0}^{p}(50),$ i.e. $%
\lambda _{c}(50,500)\thickapprox \lambda _{c}(50,0),$ and the improvement of
the $\mathcal{L}_{l}^{p}(50)$ network performance is observed for $l$
greater than $500.$

This rather unexpected phenomenon can be understood as follows. When
additional links are introduced, and their number is less than $l_{c}(m)$,
they provide a shortcut between distant parts of the network. Since packets
are forwarded to their destinations via the shortest path, it often happens
that one link serves as a shortcut for many packets from the neighborhood.
One could say that additional links ``attract'' most of the traffic and
quickly become congested, even though sites which are not close to extra
links are almost empty. This is well illustrated in Figure~\ref{fig700},
which shows snapshots of dynamics of the network with $R_{\infty }$ routing
algorithm, $50\times 50$ nodes and periodic boundary conditions. The
presented load is $\lambda (50,0)=0.025$, just below the critical value $%
\lambda _{c}(50,0)=0.028$. If there are no additional random links the
network dynamics remains in the steady state, as is showed by the left
column of Figure \ref{fig700}. The number of packets in the network
fluctuates slightly over time, but remains at the same level: at $k=100$
there are $2106$ packets in the network, while at $k=1000$ there are $2127$
packets. When additional $100$ random links are introduced, which is less
than $l_{c}(50)$, keeping all other network parameters unchanged, the
network dynamics enters the congested phase, as it is illustrated by the
right column of Figure \ref{fig700}. The number of packets in the network
increases rapidly over time from $2238$ packets in the network at $k=100$,
to $14990$ packets at $k=1000$. Congestions occur mainly at inputs and
exists from the extra links. At these nodes the queue sizes are
substantially larger than in other nodes of the network. This is illustrated
by the dark spots in the figures of the right column of Figure~\ref{fig700}.

\subsection{Partial table routing}

The performance of a network changes from the one described above when the
value of the cutoff parameter $m$ is less than $D_{max},$ i.e. $m<D_{max}.$
For\ the routing algorithm $R_{20}$ applied on the lattice $\mathcal{L}%
_{l}^{p}(50)$ for various values of the parameter $l$ the performance of the
network is shown on Figure~\ref{fig600}d. From this figure we observe that
the critical value $l_{c}(m)$ below which the performance of the network
with the routing algorithm $R_{20}$ is worse than the performance of the
network without random links added is rather low. This value is lower than
the corresponding value when $R_{\infty }$ routing algorithm has been used.
For example, adding more than about $15$ links increases the critical load $%
\lambda _{c}$. Adding about $50$ links, just a $2\%$ increment in the number
of links, increases the critical load by $100\%$ ! The performance of the
network improves significantly further with the increase of the number of
random links.

The explanation of this behavior is straightforward. As we have already
mentioned, for a regular square lattice without random links added, the
values of the critical load $\lambda _{c}(m,0)$ strongly decrease with
decrease of the cutoff parameter $m$. This results from the fact that
packets which are further away than $m$ units from their destinations are
not being attracted to the destinations and travel randomly through the
network. However, addition of random links significantly decreases the
average distances between network nodes in the metric $d_{SP}$. Therefore,
it does not matter what is the exact value of the cutoff parameter $m$.
Most of the time distances in the metric $d_{SP}$ are way below $m$ and
packets can be attracted to their destinations much faster. This attraction
increases with the increase in the number of additional random links. Hence,
when additional links are present, the critical load $\lambda _{c}(m,l)$ is
not very much dependent on $m$, unless $m$ is very small and increases with
increase in value of the number $l$ of the random links added.

Let $\lambda _{c}(m,l)$ be the critical load of a network with the
cutoff parameter $m$ and with $l$ extra links added. Let
$m_{1}=D_{max}$ and let $m_{2}$ be smaller than $m_{1}$, but not too
close to $1,$ for example, $m_{1}=50$, $m_{2}=20$, as in
Figure~\ref{fig600}. The performance of the networks with the routing
algorithms $R_{m_{1}}$ and $R_{m_{2}}$, as in the Figure
\ref{fig600}, can be summarized as follows
\begin{eqnarray*}
l_{c}(m_{2})&<&l_{c}(m_{1}),\\
\lambda _{c}(m_{2},0)&<&\lambda _{c}(m_{1},0),\\
\lambda _{c}(m_{i},l)&<&\lambda _{c}(m_{i},0),
\end{eqnarray*}
for $i=1,2$ when $l<l_{c}(m_{i})$ and
\[
\lambda _{c}(m_{i},0)<\lambda _{c}(m_{i},l),
\]
for $i=1,2$ when $l>l_{c}(m_{i})$, and for sufficiently large $l$
\[
\lambda _{c}(m_{1},l)\approx \lambda _{c}(m_{2},l).
\]

Figure~\ref{fig800} shows how the introduction of $100$ additional links to
a network with $50\times 50$ nodes, routing algorithm $R_{20}$ and presented
load $\lambda =0.008$ (which is above the critical load for $l=0$), affects
the network dynamics. When $l=0$ , the number of packets in the network
grows with time, from $1248$ at $k=100$ to $6750$ at $k=1000$, indicating
that the system is in the congested state. The left column of Figure \ref
{fig800} shows that the queue sizes grow almost uniformly over all nodes of
the network. Hence, the congestion is distributed uniformly over all nodes
of the network. However, when $l=100$ additional random links are
introduced, the right column of Figure \ref{fig800} shows that congestion is
eliminated. The number of packets in the network remains almost steady and
it fluctuates around 230 (Figure~\ref{fig800}). If an occasional small
congestion occurs near the entrance to one of the shortcuts, it quickly
disappears. For example, dark square visible in the right column of Figure~%
\ref{fig800} at $k=100$ is not visible at $k=1000$, demonstrating that local
congestions are not permanent.

\section{Conclusion}

We found that the impact of additional random links on the performance of
the network strongly depends on the routing scheme used in the network.
Critical load of a network can be notably improved if the routing is based
on a ``geometrical distance''. Adding small number of additional links can
decrease the average delay and shift the transition to the jammed phase
toward higher load values. This, in general, is not true for routing schemes
based on the ``shortest path metric''. In this case, if the number of
additional links is small, one can actually observe degradation of
performance: many packets attempt to utilize shortcuts introduced by
additional links, causing congestion which in effect pushes the network to a
jammed phase. Reduced table routing can, to some extent, eliminate this
problem. If packets located further than $m$ links from the destination are
routed randomly (other packets taking the shortest possible path),
performance gains obtained by adding sufficient number of extra links can be
quite significant.

In order to relate our findings to data network protocols used in practice,
more research is clearly needed. In particular, congestion control
mechanisms built into protocols such as TCP/IP will certainly affect
phenomena reported here. This issue, as well as other possible modifications
of the model, is currently under investigation. Furthermore, the authors
believe that some of the issues raised in \cite{taqqu}, related to
self-similar traffic modeling and analysis, and performance modeling of
modern high-speed networks can be addressed by the methodology of this
paper.\\

\noindent \textbf{Acknowledgment}
The authors acknowledge partial financial support from the Natural Sciences
and Engineering Research Council (NSERC) of Canada and The Fields Institute
for Research in Mathematical Sciences. They express their gratitude to Bruno
Di Stefano and Murad S. Taqqu for helpful discussions.

\begin{figure}[h]
\begin{center}
\setlength{\unitlength}{0.240900pt}
\begin{picture}(1500,900)(0,0)
\tenrm
\thicklines \path(219,134)(239,134)
\thicklines \path(1436,134)(1416,134)
\put(197,134){\makebox(0,0)[r]{0}}
\thicklines \path(219,254)(239,254)
\thicklines \path(1436,254)(1416,254)
\put(197,254){\makebox(0,0)[r]{200}}
\thicklines \path(219,374)(239,374)
\thicklines \path(1436,374)(1416,374)
\put(197,374){\makebox(0,0)[r]{400}}
\thicklines \path(219,494)(239,494)
\thicklines \path(1436,494)(1416,494)
\put(197,494){\makebox(0,0)[r]{600}}
\thicklines \path(219,615)(239,615)
\thicklines \path(1436,615)(1416,615)
\put(197,615){\makebox(0,0)[r]{800}}
\thicklines \path(219,735)(239,735)
\thicklines \path(1436,735)(1416,735)
\put(197,735){\makebox(0,0)[r]{1000}}
\thicklines \path(219,855)(239,855)
\thicklines \path(1436,855)(1416,855)
\put(197,855){\makebox(0,0)[r]{1200}}
\thicklines \path(219,134)(219,154)
\thicklines \path(219,855)(219,835)
\put(219,89){\makebox(0,0){0}}
\thicklines \path(371,134)(371,154)
\thicklines \path(371,855)(371,835)
\put(371,89){\makebox(0,0){0.01}}
\thicklines \path(523,134)(523,154)
\thicklines \path(523,855)(523,835)
\put(523,89){\makebox(0,0){0.02}}
\thicklines \path(675,134)(675,154)
\thicklines \path(675,855)(675,835)
\put(675,89){\makebox(0,0){0.03}}
\thicklines \path(828,134)(828,154)
\thicklines \path(828,855)(828,835)
\put(828,89){\makebox(0,0){0.04}}
\thicklines \path(980,134)(980,154)
\thicklines \path(980,855)(980,835)
\put(980,89){\makebox(0,0){0.05}}
\thicklines \path(1132,134)(1132,154)
\thicklines \path(1132,855)(1132,835)
\put(1132,89){\makebox(0,0){0.06}}
\thicklines \path(1284,134)(1284,154)
\thicklines \path(1284,855)(1284,835)
\put(1284,89){\makebox(0,0){0.07}}
\thicklines \path(1436,134)(1436,154)
\thicklines \path(1436,855)(1436,835)
\put(1436,89){\makebox(0,0){0.08}}
\thicklines \path(219,134)(1436,134)(1436,855)(219,855)(219,134)
\put(45,494){\makebox(0,0)[l]{\shortstack{${\overline{\tau}}$}}}
\put(827,21){\makebox(0,0){${\lambda}$}}
\put(341,711){\makebox(0,0)[l]{(a)}}
\thinlines \path(234,149)(234,149)(265,150)(295,150)(325,150)(356,151)(386,151)(417,152)(447,152)(478,152)(508,153)(538,154)(569,154)(599,155)(630,156)(660,157)(691,158)(721,160)(751,162)(782,166)(812,174)(843,192)(873,208)(904,236)(934,256)(964,273)(995,291)(1025,308)(1056,320)(1086,332)(1117,344)(1147,354)(1177,364)(1208,369)(1238,377)(1269,384)(1299,392)(1330,397)(1360,403)(1390,405)(1421,410)
\put(234,149){\circle{18}}
\put(265,150){\circle{18}}
\put(295,150){\circle{18}}
\put(325,150){\circle{18}}
\put(356,151){\circle{18}}
\put(386,151){\circle{18}}
\put(417,152){\circle{18}}
\put(447,152){\circle{18}}
\put(478,152){\circle{18}}
\put(508,153){\circle{18}}
\put(538,154){\circle{18}}
\put(569,154){\circle{18}}
\put(599,155){\circle{18}}
\put(630,156){\circle{18}}
\put(660,157){\circle{18}}
\put(691,158){\circle{18}}
\put(721,160){\circle{18}}
\put(751,162){\circle{18}}
\put(782,166){\circle{18}}
\put(812,174){\circle{18}}
\put(843,192){\circle{18}}
\put(873,208){\circle{18}}
\put(904,236){\circle{18}}
\put(934,256){\circle{18}}
\put(964,273){\circle{18}}
\put(995,291){\circle{18}}
\put(1025,308){\circle{18}}
\put(1056,320){\circle{18}}
\put(1086,332){\circle{18}}
\put(1117,344){\circle{18}}
\put(1147,354){\circle{18}}
\put(1177,364){\circle{18}}
\put(1208,369){\circle{18}}
\put(1238,377){\circle{18}}
\put(1269,384){\circle{18}}
\put(1299,392){\circle{18}}
\put(1330,397){\circle{18}}
\put(1360,403){\circle{18}}
\put(1390,405){\circle{18}}
\put(1421,410){\circle{18}}
\thinlines \path(234,149)(234,149)(265,150)(295,150)(325,150)(356,151)(386,151)(417,151)(447,152)(478,152)(508,153)(538,154)(569,154)(599,155)(630,156)(660,157)(691,158)(721,160)(751,162)(782,167)(812,175)(843,205)(873,248)(904,286)(934,323)(964,361)(995,385)(1025,412)(1056,435)(1086,459)(1117,476)(1147,490)(1177,508)(1208,526)(1238,534)(1269,545)(1299,555)(1330,567)(1360,575)(1390,583)(1421,592)
\put(234,149){\circle*{18}}
\put(265,150){\circle*{18}}
\put(295,150){\circle*{18}}
\put(325,150){\circle*{18}}
\put(356,151){\circle*{18}}
\put(386,151){\circle*{18}}
\put(417,151){\circle*{18}}
\put(447,152){\circle*{18}}
\put(478,152){\circle*{18}}
\put(508,153){\circle*{18}}
\put(538,154){\circle*{18}}
\put(569,154){\circle*{18}}
\put(599,155){\circle*{18}}
\put(630,156){\circle*{18}}
\put(660,157){\circle*{18}}
\put(691,158){\circle*{18}}
\put(721,160){\circle*{18}}
\put(751,162){\circle*{18}}
\put(782,167){\circle*{18}}
\put(812,175){\circle*{18}}
\put(843,205){\circle*{18}}
\put(873,248){\circle*{18}}
\put(904,286){\circle*{18}}
\put(934,323){\circle*{18}}
\put(964,361){\circle*{18}}
\put(995,385){\circle*{18}}
\put(1025,412){\circle*{18}}
\put(1056,435){\circle*{18}}
\put(1086,459){\circle*{18}}
\put(1117,476){\circle*{18}}
\put(1147,490){\circle*{18}}
\put(1177,508){\circle*{18}}
\put(1208,526){\circle*{18}}
\put(1238,534){\circle*{18}}
\put(1269,545){\circle*{18}}
\put(1299,555){\circle*{18}}
\put(1330,567){\circle*{18}}
\put(1360,575){\circle*{18}}
\put(1390,583){\circle*{18}}
\put(1421,592){\circle*{18}}
\thinlines \path(234,149)(234,149)(265,150)(295,150)(325,150)(356,151)(386,151)(417,151)(447,152)(478,152)(508,153)(538,154)(569,154)(599,155)(630,156)(660,157)(691,158)(721,160)(751,162)(782,167)(812,178)(843,217)(873,278)(904,337)(934,396)(964,441)(995,482)(1025,517)(1056,552)(1086,585)(1117,609)(1147,633)(1177,654)(1208,676)(1238,691)(1269,708)(1299,725)(1330,737)(1360,749)(1390,762)(1421,772)
\put(234,149){\makebox(0,0){$\star$}}
\put(265,150){\makebox(0,0){$\star$}}
\put(295,150){\makebox(0,0){$\star$}}
\put(325,150){\makebox(0,0){$\star$}}
\put(356,151){\makebox(0,0){$\star$}}
\put(386,151){\makebox(0,0){$\star$}}
\put(417,151){\makebox(0,0){$\star$}}
\put(447,152){\makebox(0,0){$\star$}}
\put(478,152){\makebox(0,0){$\star$}}
\put(508,153){\makebox(0,0){$\star$}}
\put(538,154){\makebox(0,0){$\star$}}
\put(569,154){\makebox(0,0){$\star$}}
\put(599,155){\makebox(0,0){$\star$}}
\put(630,156){\makebox(0,0){$\star$}}
\put(660,157){\makebox(0,0){$\star$}}
\put(691,158){\makebox(0,0){$\star$}}
\put(721,160){\makebox(0,0){$\star$}}
\put(751,162){\makebox(0,0){$\star$}}
\put(782,167){\makebox(0,0){$\star$}}
\put(812,178){\makebox(0,0){$\star$}}
\put(843,217){\makebox(0,0){$\star$}}
\put(873,278){\makebox(0,0){$\star$}}
\put(904,337){\makebox(0,0){$\star$}}
\put(934,396){\makebox(0,0){$\star$}}
\put(964,441){\makebox(0,0){$\star$}}
\put(995,482){\makebox(0,0){$\star$}}
\put(1025,517){\makebox(0,0){$\star$}}
\put(1056,552){\makebox(0,0){$\star$}}
\put(1086,585){\makebox(0,0){$\star$}}
\put(1117,609){\makebox(0,0){$\star$}}
\put(1147,633){\makebox(0,0){$\star$}}
\put(1177,654){\makebox(0,0){$\star$}}
\put(1208,676){\makebox(0,0){$\star$}}
\put(1238,691){\makebox(0,0){$\star$}}
\put(1269,708){\makebox(0,0){$\star$}}
\put(1299,725){\makebox(0,0){$\star$}}
\put(1330,737){\makebox(0,0){$\star$}}
\put(1360,749){\makebox(0,0){$\star$}}
\put(1390,762){\makebox(0,0){$\star$}}
\put(1421,772){\makebox(0,0){$\star$}}
\end{picture} 
\setlength{\unitlength}{0.240900pt}
\begin{picture}(1500,900)(0,0)
\tenrm
\thicklines \path(197,134)(217,134)
\thicklines \path(1436,134)(1416,134)
\put(175,134){\makebox(0,0)[r]{0}}
\thicklines \path(197,254)(217,254)
\thicklines \path(1436,254)(1416,254)
\put(175,254){\makebox(0,0)[r]{20}}
\thicklines \path(197,374)(217,374)
\thicklines \path(1436,374)(1416,374)
\put(175,374){\makebox(0,0)[r]{40}}
\thicklines \path(197,495)(217,495)
\thicklines \path(1436,495)(1416,495)
\put(175,495){\makebox(0,0)[r]{60}}
\thicklines \path(197,615)(217,615)
\thicklines \path(1436,615)(1416,615)
\put(175,615){\makebox(0,0)[r]{80}}
\thicklines \path(197,735)(217,735)
\thicklines \path(1436,735)(1416,735)
\put(175,735){\makebox(0,0)[r]{100}}
\thicklines \path(197,855)(217,855)
\thicklines \path(1436,855)(1416,855)
\put(175,855){\makebox(0,0)[r]{120}}
\thicklines \path(197,134)(197,154)
\thicklines \path(197,855)(197,835)
\put(197,89){\makebox(0,0){0}}
\thicklines \path(352,134)(352,154)
\thicklines \path(352,855)(352,835)
\put(352,89){\makebox(0,0){0.01}}
\thicklines \path(507,134)(507,154)
\thicklines \path(507,855)(507,835)
\put(507,89){\makebox(0,0){0.02}}
\thicklines \path(662,134)(662,154)
\thicklines \path(662,855)(662,835)
\put(662,89){\makebox(0,0){0.03}}
\thicklines \path(817,134)(817,154)
\thicklines \path(817,855)(817,835)
\put(817,89){\makebox(0,0){0.04}}
\thicklines \path(971,134)(971,154)
\thicklines \path(971,855)(971,835)
\put(971,89){\makebox(0,0){0.05}}
\thicklines \path(1126,134)(1126,154)
\thicklines \path(1126,855)(1126,835)
\put(1126,89){\makebox(0,0){0.06}}
\thicklines \path(1281,134)(1281,154)
\thicklines \path(1281,855)(1281,835)
\put(1281,89){\makebox(0,0){0.07}}
\thicklines \path(1436,134)(1436,154)
\thicklines \path(1436,855)(1436,835)
\put(1436,89){\makebox(0,0){0.08}}
\thicklines \path(197,134)(1436,134)(1436,855)(197,855)(197,134)
\put(45,494){\makebox(0,0)[l]{\shortstack{$\displaystyle{{\frac{N}{L^2}}}$}}}
\put(816,21){\makebox(0,0){${\lambda}$}}
\put(321,711){\makebox(0,0)[l]{(b)}}
\thinlines \path(212,134)(212,134)(243,134)(274,135)(305,135)(336,135)(367,136)(398,136)(429,137)(460,137)(491,137)(522,138)(553,139)(584,139)(615,140)(646,141)(677,141)(708,142)(739,144)(770,146)(801,150)(832,159)(863,173)(894,193)(925,210)(956,227)(987,245)(1018,264)(1049,279)(1080,296)(1111,312)(1142,328)(1173,346)(1204,359)(1235,378)(1266,392)(1297,406)(1328,423)(1359,440)(1390,452)(1421,468)
\put(212,134){\circle{18}}
\put(243,134){\circle{18}}
\put(274,135){\circle{18}}
\put(305,135){\circle{18}}
\put(336,135){\circle{18}}
\put(367,136){\circle{18}}
\put(398,136){\circle{18}}
\put(429,137){\circle{18}}
\put(460,137){\circle{18}}
\put(491,137){\circle{18}}
\put(522,138){\circle{18}}
\put(553,139){\circle{18}}
\put(584,139){\circle{18}}
\put(615,140){\circle{18}}
\put(646,141){\circle{18}}
\put(677,141){\circle{18}}
\put(708,142){\circle{18}}
\put(739,144){\circle{18}}
\put(770,146){\circle{18}}
\put(801,150){\circle{18}}
\put(832,159){\circle{18}}
\put(863,173){\circle{18}}
\put(894,193){\circle{18}}
\put(925,210){\circle{18}}
\put(956,227){\circle{18}}
\put(987,245){\circle{18}}
\put(1018,264){\circle{18}}
\put(1049,279){\circle{18}}
\put(1080,296){\circle{18}}
\put(1111,312){\circle{18}}
\put(1142,328){\circle{18}}
\put(1173,346){\circle{18}}
\put(1204,359){\circle{18}}
\put(1235,378){\circle{18}}
\put(1266,392){\circle{18}}
\put(1297,406){\circle{18}}
\put(1328,423){\circle{18}}
\put(1359,440){\circle{18}}
\put(1390,452){\circle{18}}
\put(1421,468){\circle{18}}
\thinlines \path(212,134)(212,134)(243,134)(274,135)(305,135)(336,135)(367,136)(398,136)(429,137)(460,137)(491,137)(522,138)(553,139)(584,139)(615,140)(646,140)(677,142)(708,142)(739,144)(770,146)(801,148)(832,165)(863,190)(894,216)(925,245)(956,274)(987,295)(1018,321)(1049,347)(1080,375)(1111,398)(1142,421)(1173,450)(1204,473)(1235,493)(1266,516)(1297,541)(1328,566)(1359,586)(1390,611)(1421,633)
\put(212,134){\circle*{18}}
\put(243,134){\circle*{18}}
\put(274,135){\circle*{18}}
\put(305,135){\circle*{18}}
\put(336,135){\circle*{18}}
\put(367,136){\circle*{18}}
\put(398,136){\circle*{18}}
\put(429,137){\circle*{18}}
\put(460,137){\circle*{18}}
\put(491,137){\circle*{18}}
\put(522,138){\circle*{18}}
\put(553,139){\circle*{18}}
\put(584,139){\circle*{18}}
\put(615,140){\circle*{18}}
\put(646,140){\circle*{18}}
\put(677,142){\circle*{18}}
\put(708,142){\circle*{18}}
\put(739,144){\circle*{18}}
\put(770,146){\circle*{18}}
\put(801,148){\circle*{18}}
\put(832,165){\circle*{18}}
\put(863,190){\circle*{18}}
\put(894,216){\circle*{18}}
\put(925,245){\circle*{18}}
\put(956,274){\circle*{18}}
\put(987,295){\circle*{18}}
\put(1018,321){\circle*{18}}
\put(1049,347){\circle*{18}}
\put(1080,375){\circle*{18}}
\put(1111,398){\circle*{18}}
\put(1142,421){\circle*{18}}
\put(1173,450){\circle*{18}}
\put(1204,473){\circle*{18}}
\put(1235,493){\circle*{18}}
\put(1266,516){\circle*{18}}
\put(1297,541){\circle*{18}}
\put(1328,566){\circle*{18}}
\put(1359,586){\circle*{18}}
\put(1390,611){\circle*{18}}
\put(1421,633){\circle*{18}}
\thinlines \path(212,134)(212,134)(243,134)(274,135)(305,135)(336,136)(367,136)(398,136)(429,137)(460,137)(491,137)(522,138)(553,139)(584,139)(615,140)(646,141)(677,141)(708,142)(739,144)(770,147)(801,150)(832,171)(863,204)(894,239)(925,279)(956,314)(987,347)(1018,381)(1049,417)(1080,453)(1111,484)(1142,518)(1173,549)(1204,582)(1235,613)(1266,642)(1297,676)(1328,709)(1359,738)(1390,770)(1421,796)
\put(212,134){\makebox(0,0){$\star$}}
\put(243,134){\makebox(0,0){$\star$}}
\put(274,135){\makebox(0,0){$\star$}}
\put(305,135){\makebox(0,0){$\star$}}
\put(336,136){\makebox(0,0){$\star$}}
\put(367,136){\makebox(0,0){$\star$}}
\put(398,136){\makebox(0,0){$\star$}}
\put(429,137){\makebox(0,0){$\star$}}
\put(460,137){\makebox(0,0){$\star$}}
\put(491,137){\makebox(0,0){$\star$}}
\put(522,138){\makebox(0,0){$\star$}}
\put(553,139){\makebox(0,0){$\star$}}
\put(584,139){\makebox(0,0){$\star$}}
\put(615,140){\makebox(0,0){$\star$}}
\put(646,141){\makebox(0,0){$\star$}}
\put(677,141){\makebox(0,0){$\star$}}
\put(708,142){\makebox(0,0){$\star$}}
\put(739,144){\makebox(0,0){$\star$}}
\put(770,147){\makebox(0,0){$\star$}}
\put(801,150){\makebox(0,0){$\star$}}
\put(832,171){\makebox(0,0){$\star$}}
\put(863,204){\makebox(0,0){$\star$}}
\put(894,239){\makebox(0,0){$\star$}}
\put(925,279){\makebox(0,0){$\star$}}
\put(956,314){\makebox(0,0){$\star$}}
\put(987,347){\makebox(0,0){$\star$}}
\put(1018,381){\makebox(0,0){$\star$}}
\put(1049,417){\makebox(0,0){$\star$}}
\put(1080,453){\makebox(0,0){$\star$}}
\put(1111,484){\makebox(0,0){$\star$}}
\put(1142,518){\makebox(0,0){$\star$}}
\put(1173,549){\makebox(0,0){$\star$}}
\put(1204,582){\makebox(0,0){$\star$}}
\put(1235,613){\makebox(0,0){$\star$}}
\put(1266,642){\makebox(0,0){$\star$}}
\put(1297,676){\makebox(0,0){$\star$}}
\put(1328,709){\makebox(0,0){$\star$}}
\put(1359,738){\makebox(0,0){$\star$}}
\put(1390,770){\makebox(0,0){$\star$}}
\put(1421,796){\makebox(0,0){$\star$}}
\end{picture}
\end{center}
\caption{(a) Average lifetime of a packet $\overline{\protect\tau}(k)$ as a
function of $\protect\lambda$, for $\mathcal{L}_0^p(50)$ with $d_{PM}$ and $%
m=D_{max}$ after $k=1000$ ($\circ$), $k=1500$ ($\bullet$), and $k=2000$ ($%
\star$) iterations. (b) Number of packets in the system $N(k)$ as a function
of $\protect\lambda$ after $k=1000$ ($\circ$), $k=1500$ ($\bullet$), and $%
k=2000$ ($\star$) iterations. }
\label{fig100}
\end{figure}

\begin{figure}[h]
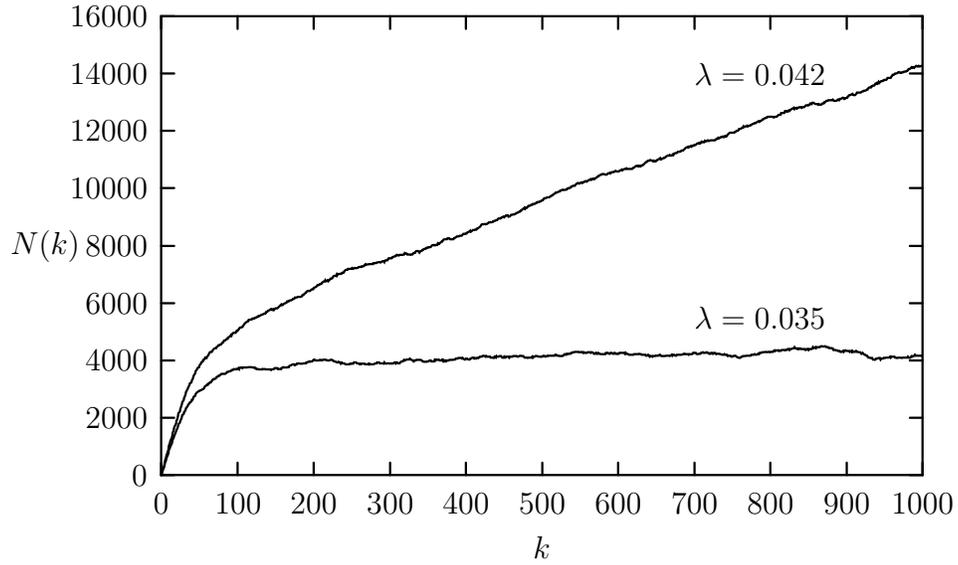

\begin{center}
\include{fig900}
\end{center}
\caption{Number of packets in the system $N(k)$ for subcritical and
supercritical values of $\protect\lambda$ ($\protect\lambda=0.035$ and $%
\protect\lambda=0.042$, respectively). $\mathcal{L}^{p}_0(50)$ with $d_{PM}$
metric and $m=D_{max}$.}
\label{fig900}
\end{figure}

\begin{figure}[h]
\begin{center}
\setlength{\unitlength}{0.240900pt}
\begin{picture}(1500,900)(0,0)
\tenrm
\thicklines \path(219,134)(239,134)
\thicklines \path(1436,134)(1416,134)
\put(197,134){\makebox(0,0)[r]{0}}
\thicklines \path(219,224)(239,224)
\thicklines \path(1436,224)(1416,224)
\put(197,224){\makebox(0,0)[r]{0.01}}
\thicklines \path(219,314)(239,314)
\thicklines \path(1436,314)(1416,314)
\put(197,314){\makebox(0,0)[r]{0.02}}
\thicklines \path(219,404)(239,404)
\thicklines \path(1436,404)(1416,404)
\put(197,404){\makebox(0,0)[r]{0.03}}
\thicklines \path(219,495)(239,495)
\thicklines \path(1436,495)(1416,495)
\put(197,495){\makebox(0,0)[r]{0.04}}
\thicklines \path(219,585)(239,585)
\thicklines \path(1436,585)(1416,585)
\put(197,585){\makebox(0,0)[r]{0.05}}
\thicklines \path(219,675)(239,675)
\thicklines \path(1436,675)(1416,675)
\put(197,675){\makebox(0,0)[r]{0.06}}
\thicklines \path(219,765)(239,765)
\thicklines \path(1436,765)(1416,765)
\put(197,765){\makebox(0,0)[r]{0.07}}
\thicklines \path(219,855)(239,855)
\thicklines \path(1436,855)(1416,855)
\put(197,855){\makebox(0,0)[r]{0.08}}
\thicklines \path(219,134)(219,154)
\thicklines \path(219,855)(219,835)
\put(219,89){\makebox(0,0){0}}
\thicklines \path(422,134)(422,154)
\thicklines \path(422,855)(422,835)
\put(422,89){\makebox(0,0){0.01}}
\thicklines \path(625,134)(625,154)
\thicklines \path(625,855)(625,835)
\put(625,89){\makebox(0,0){0.02}}
\thicklines \path(828,134)(828,154)
\thicklines \path(828,855)(828,835)
\put(828,89){\makebox(0,0){0.03}}
\thicklines \path(1030,134)(1030,154)
\thicklines \path(1030,855)(1030,835)
\put(1030,89){\makebox(0,0){0.04}}
\thicklines \path(1233,134)(1233,154)
\thicklines \path(1233,855)(1233,835)
\put(1233,89){\makebox(0,0){0.05}}
\thicklines \path(1436,134)(1436,154)
\thicklines \path(1436,855)(1436,835)
\put(1436,89){\makebox(0,0){0.06}}
\thicklines \path(219,134)(1436,134)(1436,855)(219,855)(219,134)
\put(13,494){\makebox(0,0)[l]{\shortstack{${\displaystyle \frac{N}{\overline{\tau} L^2}}$}}}
\put(827,44){\makebox(0,0){${\lambda}$}}
\put(239,142){\circle*{18}}
\put(280,162){\circle*{18}}
\put(320,182){\circle*{18}}
\put(361,190){\circle*{18}}
\put(402,219){\circle*{18}}
\put(442,230){\circle*{18}}
\put(483,248){\circle*{18}}
\put(523,262){\circle*{18}}
\put(564,281){\circle*{18}}
\put(604,302){\circle*{18}}
\put(645,320){\circle*{18}}
\put(686,340){\circle*{18}}
\put(726,352){\circle*{18}}
\put(767,370){\circle*{18}}
\put(807,385){\circle*{18}}
\put(848,412){\circle*{18}}
\put(888,418){\circle*{18}}
\put(929,440){\circle*{18}}
\put(969,453){\circle*{18}}
\put(1010,488){\circle*{18}}
\put(1051,542){\circle*{18}}
\put(1091,574){\circle*{18}}
\put(1132,623){\circle*{18}}
\put(1172,655){\circle*{18}}
\put(1213,679){\circle*{18}}
\put(1253,717){\circle*{18}}
\put(1294,747){\circle*{18}}
\put(1335,772){\circle*{18}}
\put(1375,801){\circle*{18}}
\put(1416,828){\circle*{18}}
\thicklines \path(219,134)(219,134)(225,137)(231,139)(237,142)(243,145)(250,148)(256,150)(262,153)(268,156)(274,158)(280,161)(286,164)(292,167)(299,169)(305,172)(311,175)(317,177)(323,180)(329,183)(335,186)(341,188)(347,191)(354,194)(360,196)(366,199)(372,202)(378,205)(384,207)(390,210)(396,213)(402,216)(409,218)(415,221)(421,224)(427,226)(433,229)(439,232)(445,235)(451,237)(458,240)(464,243)(470,245)(476,248)(482,251)(488,254)(494,256)(500,259)(506,262)(513,264)(519,267)
\thicklines \path(519,267)(525,270)(531,273)(537,275)(543,278)(549,281)(555,283)(561,286)(568,289)(574,292)(580,294)(586,297)(592,300)(598,302)(604,305)(610,308)(617,311)(623,313)(629,316)(635,319)(641,321)(647,324)(653,327)(659,330)(665,332)(672,335)(678,338)(684,341)(690,343)(696,346)(702,349)(708,351)(714,354)(720,357)(727,360)(733,362)(739,365)(745,368)(751,370)(757,373)(763,376)(769,379)(776,381)(782,384)(788,387)(794,389)(800,392)(806,395)(812,398)(818,400)(824,403)
\thicklines \path(824,403)(831,406)(837,408)(843,411)(849,414)(855,417)(861,419)(867,422)(873,425)(879,427)(886,430)(892,433)(898,436)(904,438)(910,441)(916,444)(922,446)(928,449)(935,452)(941,455)(947,457)(953,460)(959,463)(965,466)(971,468)(977,471)(983,474)(990,476)(996,479)(1002,482)(1008,485)(1014,487)(1020,490)(1026,493)(1032,495)(1038,498)(1045,501)(1051,504)(1057,506)(1063,509)(1069,512)(1075,514)(1081,517)(1087,520)(1094,523)(1100,525)(1106,528)(1112,531)(1118,533)(1124,536)(1130,539)
\thicklines \path(1130,539)(1136,542)(1142,544)(1149,547)(1155,550)(1161,552)(1167,555)(1173,558)(1179,561)(1185,563)(1191,566)(1197,569)(1204,571)(1210,574)(1216,577)(1222,580)(1228,582)(1234,585)(1240,588)(1246,591)(1253,593)(1259,596)(1265,599)(1271,601)(1277,604)(1283,607)(1289,610)(1295,612)(1301,615)(1308,618)(1314,620)(1320,623)(1326,626)(1332,629)(1338,631)(1344,634)(1350,637)(1356,639)(1363,642)(1369,645)(1375,648)(1381,650)(1387,653)(1393,656)(1399,658)(1405,661)(1412,664)(1418,667)(1424,669)(1430,672)(1436,675)
\end{picture}
\end{center}
\caption{Verification of Little's law for a lattice $\mathcal{L}_0^p(50)$
with $d_{PM}$ and $m=D_{max}$, at $k=1500$. Continuous line
corresponds to $
L^2)=\protect\lambda$.}
\label{fig300}
\end{figure}
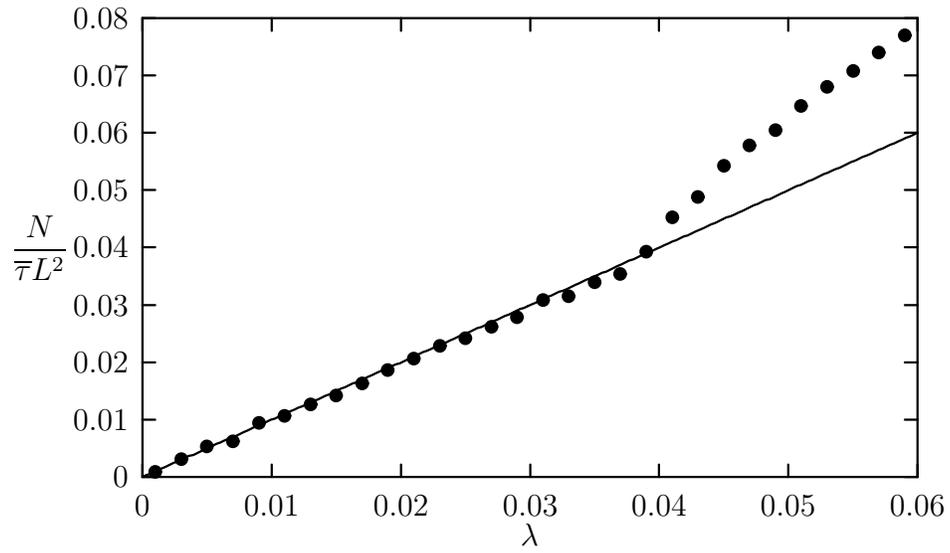

\begin{figure}[h]
\begin{center}
\setlength{\unitlength}{0.240900pt}
\begin{picture}(1500,900)(0,0)
\tenrm
\thicklines \path(219,134)(239,134)
\thicklines \path(1436,134)(1416,134)
\put(197,134){\makebox(0,0)[r]{0}}
\thicklines \path(219,214)(239,214)
\thicklines \path(1436,214)(1416,214)
\put(197,214){\makebox(0,0)[r]{500}}
\thicklines \path(219,294)(239,294)
\thicklines \path(1436,294)(1416,294)
\put(197,294){\makebox(0,0)[r]{1000}}
\thicklines \path(219,374)(239,374)
\thicklines \path(1436,374)(1416,374)
\put(197,374){\makebox(0,0)[r]{1500}}
\thicklines \path(219,454)(239,454)
\thicklines \path(1436,454)(1416,454)
\put(197,454){\makebox(0,0)[r]{2000}}
\thicklines \path(219,535)(239,535)
\thicklines \path(1436,535)(1416,535)
\put(197,535){\makebox(0,0)[r]{2500}}
\thicklines \path(219,615)(239,615)
\thicklines \path(1436,615)(1416,615)
\put(197,615){\makebox(0,0)[r]{3000}}
\thicklines \path(219,695)(239,695)
\thicklines \path(1436,695)(1416,695)
\put(197,695){\makebox(0,0)[r]{3500}}
\thicklines \path(219,775)(239,775)
\thicklines \path(1436,775)(1416,775)
\put(197,775){\makebox(0,0)[r]{4000}}
\thicklines \path(219,855)(239,855)
\thicklines \path(1436,855)(1416,855)
\put(197,855){\makebox(0,0)[r]{4500}}
\thicklines \path(219,134)(219,154)
\thicklines \path(219,855)(219,835)
\put(219,89){\makebox(0,0){0}}
\thicklines \path(341,134)(341,154)
\thicklines \path(341,855)(341,835)
\put(341,89){\makebox(0,0){5}}
\thicklines \path(462,134)(462,154)
\thicklines \path(462,855)(462,835)
\put(462,89){\makebox(0,0){10}}
\thicklines \path(584,134)(584,154)
\thicklines \path(584,855)(584,835)
\put(584,89){\makebox(0,0){15}}
\thicklines \path(706,134)(706,154)
\thicklines \path(706,855)(706,835)
\put(706,89){\makebox(0,0){20}}
\thicklines \path(827,134)(827,154)
\thicklines \path(827,855)(827,835)
\put(827,89){\makebox(0,0){25}}
\thicklines \path(949,134)(949,154)
\thicklines \path(949,855)(949,835)
\put(949,89){\makebox(0,0){30}}
\thicklines \path(1071,134)(1071,154)
\thicklines \path(1071,855)(1071,835)
\put(1071,89){\makebox(0,0){35}}
\thicklines \path(1193,134)(1193,154)
\thicklines \path(1193,855)(1193,835)
\put(1193,89){\makebox(0,0){40}}
\thicklines \path(1314,134)(1314,154)
\thicklines \path(1314,855)(1314,835)
\put(1314,89){\makebox(0,0){45}}
\thicklines \path(1436,134)(1436,154)
\thicklines \path(1436,855)(1436,835)
\put(1436,89){\makebox(0,0){50}}
\thicklines \path(219,134)(1436,134)(1436,855)(219,855)(219,134)
\put(13,494){\makebox(0,0)[l]{\shortstack{${\overline{\tau}_0}$}}}
\put(827,21){\makebox(0,0){${m}$}}
\thicklines \path(243,811)(243,811)(268,651)(292,558)(316,490)(341,436)(365,394)(389,359)(414,327)(438,301)(462,280)(487,260)(511,244)(535,229)(560,215)(584,204)(608,194)(633,186)(657,179)(681,173)(706,168)(730,163)(754,159)(779,156)(803,154)(827,151)(852,149)(876,148)(901,146)(925,145)(949,143)(974,142)(998,142)(1022,141)(1047,140)(1071,140)(1095,139)(1120,139)(1144,139)(1168,138)(1193,138)(1217,138)(1241,138)(1266,138)(1290,138)(1314,138)(1339,138)(1363,138)(1387,138)(1412,138)(1436,138)
\put(243,811){\circle{18}}
\put(268,651){\circle{18}}
\put(292,558){\circle{18}}
\put(316,490){\circle{18}}
\put(341,436){\circle{18}}
\put(365,394){\circle{18}}
\put(389,359){\circle{18}}
\put(414,327){\circle{18}}
\put(438,301){\circle{18}}
\put(462,280){\circle{18}}
\put(487,260){\circle{18}}
\put(511,244){\circle{18}}
\put(535,229){\circle{18}}
\put(560,215){\circle{18}}
\put(584,204){\circle{18}}
\put(608,194){\circle{18}}
\put(633,186){\circle{18}}
\put(657,179){\circle{18}}
\put(681,173){\circle{18}}
\put(706,168){\circle{18}}
\put(730,163){\circle{18}}
\put(754,159){\circle{18}}
\put(779,156){\circle{18}}
\put(803,154){\circle{18}}
\put(827,151){\circle{18}}
\put(852,149){\circle{18}}
\put(876,148){\circle{18}}
\put(901,146){\circle{18}}
\put(925,145){\circle{18}}
\put(949,143){\circle{18}}
\put(974,142){\circle{18}}
\put(998,142){\circle{18}}
\put(1022,141){\circle{18}}
\put(1047,140){\circle{18}}
\put(1071,140){\circle{18}}
\put(1095,139){\circle{18}}
\put(1120,139){\circle{18}}
\put(1144,139){\circle{18}}
\put(1168,138){\circle{18}}
\put(1193,138){\circle{18}}
\put(1217,138){\circle{18}}
\put(1241,138){\circle{18}}
\put(1266,138){\circle{18}}
\put(1290,138){\circle{18}}
\put(1314,138){\circle{18}}
\put(1339,138){\circle{18}}
\put(1363,138){\circle{18}}
\put(1387,138){\circle{18}}
\put(1412,138){\circle{18}}
\put(1436,138){\circle{18}}
\end{picture}
\end{center}
\caption{ Free packet delay $\overline{\protect\tau}_0$ as a function of $m$
for a lattice $\mathcal{L}^{p}_0(50)$ with $d_{PM}$ metric.}
\label{fig500}
\end{figure}
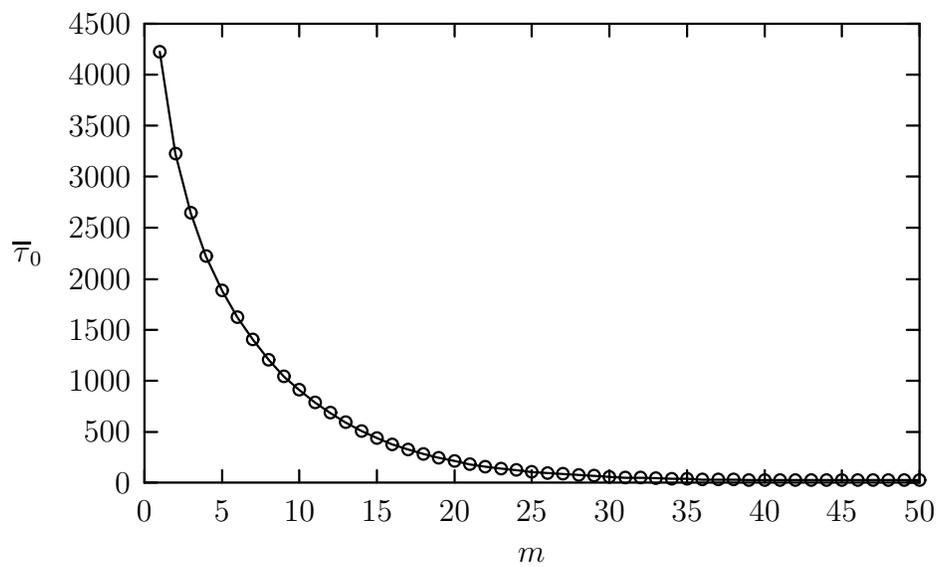

\begin{figure}[p]
\begin{center}
\setlength{\unitlength}{0.240900pt}
\begin{picture}(1500,900)(0,0)
\tenrm
\thicklines \path(197,134)(217,134)
\thicklines \path(1436,134)(1416,134)
\put(175,134){\makebox(0,0)[r]{0}}
\thicklines \path(197,237)(217,237)
\thicklines \path(1436,237)(1416,237)
\put(175,237){\makebox(0,0)[r]{100}}
\thicklines \path(197,340)(217,340)
\thicklines \path(1436,340)(1416,340)
\put(175,340){\makebox(0,0)[r]{200}}
\thicklines \path(197,443)(217,443)
\thicklines \path(1436,443)(1416,443)
\put(175,443){\makebox(0,0)[r]{300}}
\thicklines \path(197,546)(217,546)
\thicklines \path(1436,546)(1416,546)
\put(175,546){\makebox(0,0)[r]{400}}
\thicklines \path(197,649)(217,649)
\thicklines \path(1436,649)(1416,649)
\put(175,649){\makebox(0,0)[r]{500}}
\thicklines \path(197,752)(217,752)
\thicklines \path(1436,752)(1416,752)
\put(175,752){\makebox(0,0)[r]{600}}
\thicklines \path(197,855)(217,855)
\thicklines \path(1436,855)(1416,855)
\put(175,855){\makebox(0,0)[r]{700}}
\thicklines \path(197,134)(197,154)
\thicklines \path(197,855)(197,835)
\put(197,89){\makebox(0,0){0}}
\thicklines \path(404,134)(404,154)
\thicklines \path(404,855)(404,835)
\put(404,89){\makebox(0,0){0.01}}
\thicklines \path(610,134)(610,154)
\thicklines \path(610,855)(610,835)
\put(610,89){\makebox(0,0){0.02}}
\thicklines \path(816,134)(816,154)
\thicklines \path(816,855)(816,835)
\put(816,89){\makebox(0,0){0.03}}
\thicklines \path(1023,134)(1023,154)
\thicklines \path(1023,855)(1023,835)
\put(1023,89){\makebox(0,0){0.04}}
\thicklines \path(1230,134)(1230,154)
\thicklines \path(1230,855)(1230,835)
\put(1230,89){\makebox(0,0){0.05}}
\thicklines \path(1436,134)(1436,154)
\thicklines \path(1436,855)(1436,835)
\put(1436,89){\makebox(0,0){0.06}}
\thicklines \path(197,134)(1436,134)(1436,855)(197,855)(197,134)
\put(45,494){\makebox(0,0)[l]{\shortstack{${\overline{\tau}}$}}}
\put(816,21){\makebox(0,0){${\lambda}$}}
\put(321,711){\makebox(0,0)[l]{(a)}}
\thinlines \path(218,168)(218,168)(259,170)(300,172)(342,174)(383,176)(424,178)(465,181)(507,186)(548,199)(589,227)(631,276)(672,324)(713,375)(755,431)(796,476)(837,513)(878,540)(920,580)(961,616)(1002,636)(1044,660)(1085,685)(1126,707)(1168,726)(1209,746)(1250,765)(1291,779)(1333,790)(1374,804)(1415,811)(1436,821)
\put(218,168){\circle{18}}
\put(259,170){\circle{18}}
\put(300,172){\circle{18}}
\put(342,174){\circle{18}}
\put(383,176){\circle{18}}
\put(424,178){\circle{18}}
\put(465,181){\circle{18}}
\put(507,186){\circle{18}}
\put(548,199){\circle{18}}
\put(589,227){\circle{18}}
\put(631,276){\circle{18}}
\put(672,324){\circle{18}}
\put(713,375){\circle{18}}
\put(755,431){\circle{18}}
\put(796,476){\circle{18}}
\put(837,513){\circle{18}}
\put(878,540){\circle{18}}
\put(920,580){\circle{18}}
\put(961,616){\circle{18}}
\put(1002,636){\circle{18}}
\put(1044,660){\circle{18}}
\put(1085,685){\circle{18}}
\put(1126,707){\circle{18}}
\put(1168,726){\circle{18}}
\put(1209,746){\circle{18}}
\put(1250,765){\circle{18}}
\put(1291,779){\circle{18}}
\put(1333,790){\circle{18}}
\put(1374,804){\circle{18}}
\put(1415,811){\circle{18}}
\thinlines \path(218,165)(218,165)(259,166)(300,167)(342,169)(383,169)(424,171)(465,173)(507,176)(548,178)(589,182)(631,188)(672,205)(713,251)(755,291)(796,333)(837,380)(878,436)(920,486)(961,528)(1002,558)(1044,589)(1085,617)(1126,644)(1168,667)(1209,704)(1250,728)(1291,743)(1333,763)(1374,786)(1415,800)(1436,811)
\put(218,165){\circle*{18}}
\put(259,166){\circle*{18}}
\put(300,167){\circle*{18}}
\put(342,169){\circle*{18}}
\put(383,169){\circle*{18}}
\put(424,171){\circle*{18}}
\put(465,173){\circle*{18}}
\put(507,176){\circle*{18}}
\put(548,178){\circle*{18}}
\put(589,182){\circle*{18}}
\put(631,188){\circle*{18}}
\put(672,205){\circle*{18}}
\put(713,251){\circle*{18}}
\put(755,291){\circle*{18}}
\put(796,333){\circle*{18}}
\put(837,380){\circle*{18}}
\put(878,436){\circle*{18}}
\put(920,486){\circle*{18}}
\put(961,528){\circle*{18}}
\put(1002,558){\circle*{18}}
\put(1044,589){\circle*{18}}
\put(1085,617){\circle*{18}}
\put(1126,644){\circle*{18}}
\put(1168,667){\circle*{18}}
\put(1209,704){\circle*{18}}
\put(1250,728){\circle*{18}}
\put(1291,743){\circle*{18}}
\put(1333,763){\circle*{18}}
\put(1374,786){\circle*{18}}
\put(1415,800){\circle*{18}}
\thinlines \path(218,162)(218,162)(259,163)(300,164)(342,165)(383,165)(424,167)(465,168)(507,170)(548,171)(589,173)(631,177)(672,179)(713,189)(755,204)(796,254)(837,282)(878,331)(920,374)(961,424)(1002,470)(1044,509)(1085,560)(1126,584)(1168,616)(1209,647)(1250,683)(1291,703)(1333,732)(1374,755)(1415,769)(1436,775)
\put(218,162){\makebox(0,0){$\star$}}
\put(259,163){\makebox(0,0){$\star$}}
\put(300,164){\makebox(0,0){$\star$}}
\put(342,165){\makebox(0,0){$\star$}}
\put(383,165){\makebox(0,0){$\star$}}
\put(424,167){\makebox(0,0){$\star$}}
\put(465,168){\makebox(0,0){$\star$}}
\put(507,170){\makebox(0,0){$\star$}}
\put(548,171){\makebox(0,0){$\star$}}
\put(589,173){\makebox(0,0){$\star$}}
\put(631,177){\makebox(0,0){$\star$}}
\put(672,179){\makebox(0,0){$\star$}}
\put(713,189){\makebox(0,0){$\star$}}
\put(755,204){\makebox(0,0){$\star$}}
\put(796,254){\makebox(0,0){$\star$}}
\put(837,282){\makebox(0,0){$\star$}}
\put(878,331){\makebox(0,0){$\star$}}
\put(920,374){\makebox(0,0){$\star$}}
\put(961,424){\makebox(0,0){$\star$}}
\put(1002,470){\makebox(0,0){$\star$}}
\put(1044,509){\makebox(0,0){$\star$}}
\put(1085,560){\makebox(0,0){$\star$}}
\put(1126,584){\makebox(0,0){$\star$}}
\put(1168,616){\makebox(0,0){$\star$}}
\put(1209,647){\makebox(0,0){$\star$}}
\put(1250,683){\makebox(0,0){$\star$}}
\put(1291,703){\makebox(0,0){$\star$}}
\put(1333,732){\makebox(0,0){$\star$}}
\put(1374,755){\makebox(0,0){$\star$}}
\put(1415,769){\makebox(0,0){$\star$}}
\thinlines \path(218,159)(218,159)(259,159)(300,159)(342,161)(383,161)(424,162)(465,162)(507,163)(548,164)(589,166)(631,166)(672,168)(713,170)(755,173)(796,177)(837,183)(878,199)(920,223)(961,270)(1002,329)(1044,356)(1085,408)(1126,445)(1168,503)(1209,541)(1250,580)(1291,603)(1333,645)(1374,660)(1415,700)(1436,712)
\put(218,159){\makebox(0,0){$+$}}
\put(259,159){\makebox(0,0){$+$}}
\put(300,159){\makebox(0,0){$+$}}
\put(342,161){\makebox(0,0){$+$}}
\put(383,161){\makebox(0,0){$+$}}
\put(424,162){\makebox(0,0){$+$}}
\put(465,162){\makebox(0,0){$+$}}
\put(507,163){\makebox(0,0){$+$}}
\put(548,164){\makebox(0,0){$+$}}
\put(589,166){\makebox(0,0){$+$}}
\put(631,166){\makebox(0,0){$+$}}
\put(672,168){\makebox(0,0){$+$}}
\put(713,170){\makebox(0,0){$+$}}
\put(755,173){\makebox(0,0){$+$}}
\put(796,177){\makebox(0,0){$+$}}
\put(837,183){\makebox(0,0){$+$}}
\put(878,199){\makebox(0,0){$+$}}
\put(920,223){\makebox(0,0){$+$}}
\put(961,270){\makebox(0,0){$+$}}
\put(1002,329){\makebox(0,0){$+$}}
\put(1044,356){\makebox(0,0){$+$}}
\put(1085,408){\makebox(0,0){$+$}}
\put(1126,445){\makebox(0,0){$+$}}
\put(1168,503){\makebox(0,0){$+$}}
\put(1209,541){\makebox(0,0){$+$}}
\put(1250,580){\makebox(0,0){$+$}}
\put(1291,603){\makebox(0,0){$+$}}
\put(1333,645){\makebox(0,0){$+$}}
\put(1374,660){\makebox(0,0){$+$}}
\put(1415,700){\makebox(0,0){$+$}}
\end{picture} 
\setlength{\unitlength}{0.240900pt}
\begin{picture}(1500,900)(0,0)
\tenrm
\thicklines \path(197,134)(217,134)
\thicklines \path(1436,134)(1416,134)
\put(175,134){\makebox(0,0)[r]{0}}
\thicklines \path(197,224)(217,224)
\thicklines \path(1436,224)(1416,224)
\put(175,224){\makebox(0,0)[r]{100}}
\thicklines \path(197,314)(217,314)
\thicklines \path(1436,314)(1416,314)
\put(175,314){\makebox(0,0)[r]{200}}
\thicklines \path(197,404)(217,404)
\thicklines \path(1436,404)(1416,404)
\put(175,404){\makebox(0,0)[r]{300}}
\thicklines \path(197,494)(217,494)
\thicklines \path(1436,494)(1416,494)
\put(175,494){\makebox(0,0)[r]{400}}
\thicklines \path(197,585)(217,585)
\thicklines \path(1436,585)(1416,585)
\put(175,585){\makebox(0,0)[r]{500}}
\thicklines \path(197,675)(217,675)
\thicklines \path(1436,675)(1416,675)
\put(175,675){\makebox(0,0)[r]{600}}
\thicklines \path(197,765)(217,765)
\thicklines \path(1436,765)(1416,765)
\put(175,765){\makebox(0,0)[r]{700}}
\thicklines \path(197,855)(217,855)
\thicklines \path(1436,855)(1416,855)
\put(175,855){\makebox(0,0)[r]{800}}
\thicklines \path(197,134)(197,154)
\thicklines \path(197,855)(197,835)
\put(197,89){\makebox(0,0){0}}
\thicklines \path(352,134)(352,154)
\thicklines \path(352,855)(352,835)
\put(352,89){\makebox(0,0){0.01}}
\thicklines \path(507,134)(507,154)
\thicklines \path(507,855)(507,835)
\put(507,89){\makebox(0,0){0.02}}
\thicklines \path(662,134)(662,154)
\thicklines \path(662,855)(662,835)
\put(662,89){\makebox(0,0){0.03}}
\thicklines \path(817,134)(817,154)
\thicklines \path(817,855)(817,835)
\put(817,89){\makebox(0,0){0.04}}
\thicklines \path(971,134)(971,154)
\thicklines \path(971,855)(971,835)
\put(971,89){\makebox(0,0){0.05}}
\thicklines \path(1126,134)(1126,154)
\thicklines \path(1126,855)(1126,835)
\put(1126,89){\makebox(0,0){0.06}}
\thicklines \path(1281,134)(1281,154)
\thicklines \path(1281,855)(1281,835)
\put(1281,89){\makebox(0,0){0.07}}
\thicklines \path(1436,134)(1436,154)
\thicklines \path(1436,855)(1436,835)
\put(1436,89){\makebox(0,0){0.08}}
\thicklines \path(197,134)(1436,134)(1436,855)(197,855)(197,134)
\put(45,494){\makebox(0,0)[l]{\shortstack{${\overline{\tau}}$}}}
\put(816,21){\makebox(0,0){${\lambda}$}}
\put(321,711){\makebox(0,0)[l]{(b)}}
\thinlines \path(212,157)(212,157)(243,157)(274,158)(305,158)(336,159)(367,159)(398,160)(429,161)(460,161)(491,163)(522,164)(553,164)(584,166)(615,167)(646,168)(677,170)(708,173)(739,176)(770,183)(801,196)(832,240)(863,305)(894,361)(925,418)(956,475)(987,511)(1018,552)(1049,585)(1080,622)(1111,647)(1142,668)(1173,695)(1204,721)(1235,734)(1266,751)(1297,765)(1328,783)(1359,795)(1390,807)(1421,820)
\put(212,157){\circle{18}}
\put(243,157){\circle{18}}
\put(274,158){\circle{18}}
\put(305,158){\circle{18}}
\put(336,159){\circle{18}}
\put(367,159){\circle{18}}
\put(398,160){\circle{18}}
\put(429,161){\circle{18}}
\put(460,161){\circle{18}}
\put(491,163){\circle{18}}
\put(522,164){\circle{18}}
\put(553,164){\circle{18}}
\put(584,166){\circle{18}}
\put(615,167){\circle{18}}
\put(646,168){\circle{18}}
\put(677,170){\circle{18}}
\put(708,173){\circle{18}}
\put(739,176){\circle{18}}
\put(770,183){\circle{18}}
\put(801,196){\circle{18}}
\put(832,240){\circle{18}}
\put(863,305){\circle{18}}
\put(894,361){\circle{18}}
\put(925,418){\circle{18}}
\put(956,475){\circle{18}}
\put(987,511){\circle{18}}
\put(1018,552){\circle{18}}
\put(1049,585){\circle{18}}
\put(1080,622){\circle{18}}
\put(1111,647){\circle{18}}
\put(1142,668){\circle{18}}
\put(1173,695){\circle{18}}
\put(1204,721){\circle{18}}
\put(1235,734){\circle{18}}
\put(1266,751){\circle{18}}
\put(1297,765){\circle{18}}
\put(1328,783){\circle{18}}
\put(1359,795){\circle{18}}
\put(1390,807){\circle{18}}
\put(1421,820){\circle{18}}
\thinlines \path(212,155)(212,155)(243,156)(274,156)(305,157)(336,157)(367,158)(398,158)(429,159)(460,160)(491,160)(522,161)(553,162)(584,163)(615,164)(646,165)(677,166)(708,168)(739,171)(770,174)(801,179)(832,189)(863,212)(894,268)(925,322)(956,396)(987,439)(1018,481)(1049,520)(1080,552)(1111,589)(1142,623)(1173,646)(1204,671)(1235,692)(1266,707)(1297,729)(1328,747)(1359,759)(1390,778)(1421,791)
\put(212,155){\circle*{18}}
\put(243,156){\circle*{18}}
\put(274,156){\circle*{18}}
\put(305,157){\circle*{18}}
\put(336,157){\circle*{18}}
\put(367,158){\circle*{18}}
\put(398,158){\circle*{18}}
\put(429,159){\circle*{18}}
\put(460,160){\circle*{18}}
\put(491,160){\circle*{18}}
\put(522,161){\circle*{18}}
\put(553,162){\circle*{18}}
\put(584,163){\circle*{18}}
\put(615,164){\circle*{18}}
\put(646,165){\circle*{18}}
\put(677,166){\circle*{18}}
\put(708,168){\circle*{18}}
\put(739,171){\circle*{18}}
\put(770,174){\circle*{18}}
\put(801,179){\circle*{18}}
\put(832,189){\circle*{18}}
\put(863,212){\circle*{18}}
\put(894,268){\circle*{18}}
\put(925,322){\circle*{18}}
\put(956,396){\circle*{18}}
\put(987,439){\circle*{18}}
\put(1018,481){\circle*{18}}
\put(1049,520){\circle*{18}}
\put(1080,552){\circle*{18}}
\put(1111,589){\circle*{18}}
\put(1142,623){\circle*{18}}
\put(1173,646){\circle*{18}}
\put(1204,671){\circle*{18}}
\put(1235,692){\circle*{18}}
\put(1266,707){\circle*{18}}
\put(1297,729){\circle*{18}}
\put(1328,747){\circle*{18}}
\put(1359,759){\circle*{18}}
\put(1390,778){\circle*{18}}
\put(1421,791){\circle*{18}}
\thinlines \path(212,154)(212,154)(243,154)(274,155)(305,155)(336,156)(367,156)(398,157)(429,157)(460,158)(491,158)(522,159)(553,160)(584,160)(615,162)(646,162)(677,163)(708,165)(739,167)(770,170)(801,172)(832,176)(863,185)(894,208)(925,246)(956,311)(987,362)(1018,409)(1049,464)(1080,492)(1111,534)(1142,567)(1173,597)(1204,628)(1235,650)(1266,673)(1297,688)(1328,707)(1359,731)(1390,747)(1421,763)
\put(212,154){\makebox(0,0){$\star$}}
\put(243,154){\makebox(0,0){$\star$}}
\put(274,155){\makebox(0,0){$\star$}}
\put(305,155){\makebox(0,0){$\star$}}
\put(336,156){\makebox(0,0){$\star$}}
\put(367,156){\makebox(0,0){$\star$}}
\put(398,157){\makebox(0,0){$\star$}}
\put(429,157){\makebox(0,0){$\star$}}
\put(460,158){\makebox(0,0){$\star$}}
\put(491,158){\makebox(0,0){$\star$}}
\put(522,159){\makebox(0,0){$\star$}}
\put(553,160){\makebox(0,0){$\star$}}
\put(584,160){\makebox(0,0){$\star$}}
\put(615,162){\makebox(0,0){$\star$}}
\put(646,162){\makebox(0,0){$\star$}}
\put(677,163){\makebox(0,0){$\star$}}
\put(708,165){\makebox(0,0){$\star$}}
\put(739,167){\makebox(0,0){$\star$}}
\put(770,170){\makebox(0,0){$\star$}}
\put(801,172){\makebox(0,0){$\star$}}
\put(832,176){\makebox(0,0){$\star$}}
\put(863,185){\makebox(0,0){$\star$}}
\put(894,208){\makebox(0,0){$\star$}}
\put(925,246){\makebox(0,0){$\star$}}
\put(956,311){\makebox(0,0){$\star$}}
\put(987,362){\makebox(0,0){$\star$}}
\put(1018,409){\makebox(0,0){$\star$}}
\put(1049,464){\makebox(0,0){$\star$}}
\put(1080,492){\makebox(0,0){$\star$}}
\put(1111,534){\makebox(0,0){$\star$}}
\put(1142,567){\makebox(0,0){$\star$}}
\put(1173,597){\makebox(0,0){$\star$}}
\put(1204,628){\makebox(0,0){$\star$}}
\put(1235,650){\makebox(0,0){$\star$}}
\put(1266,673){\makebox(0,0){$\star$}}
\put(1297,688){\makebox(0,0){$\star$}}
\put(1328,707){\makebox(0,0){$\star$}}
\put(1359,731){\makebox(0,0){$\star$}}
\put(1390,747){\makebox(0,0){$\star$}}
\put(1421,763){\makebox(0,0){$\star$}}
\thinlines \path(212,152)(212,152)(243,152)(274,153)(305,153)(336,153)(367,154)(398,154)(429,155)(460,155)(491,156)(522,156)(553,157)(584,158)(615,158)(646,159)(677,160)(708,161)(739,162)(770,163)(801,165)(832,167)(863,169)(894,174)(925,179)(956,196)(987,239)(1018,289)(1049,348)(1080,392)(1111,437)(1142,476)(1173,510)(1204,543)(1235,576)(1266,595)(1297,618)(1328,642)(1359,672)(1390,687)(1421,710)
\put(212,152){\makebox(0,0){$+$}}
\put(243,152){\makebox(0,0){$+$}}
\put(274,153){\makebox(0,0){$+$}}
\put(305,153){\makebox(0,0){$+$}}
\put(336,153){\makebox(0,0){$+$}}
\put(367,154){\makebox(0,0){$+$}}
\put(398,154){\makebox(0,0){$+$}}
\put(429,155){\makebox(0,0){$+$}}
\put(460,155){\makebox(0,0){$+$}}
\put(491,156){\makebox(0,0){$+$}}
\put(522,156){\makebox(0,0){$+$}}
\put(553,157){\makebox(0,0){$+$}}
\put(584,158){\makebox(0,0){$+$}}
\put(615,158){\makebox(0,0){$+$}}
\put(646,159){\makebox(0,0){$+$}}
\put(677,160){\makebox(0,0){$+$}}
\put(708,161){\makebox(0,0){$+$}}
\put(739,162){\makebox(0,0){$+$}}
\put(770,163){\makebox(0,0){$+$}}
\put(801,165){\makebox(0,0){$+$}}
\put(832,167){\makebox(0,0){$+$}}
\put(863,169){\makebox(0,0){$+$}}
\put(894,174){\makebox(0,0){$+$}}
\put(925,179){\makebox(0,0){$+$}}
\put(956,196){\makebox(0,0){$+$}}
\put(987,239){\makebox(0,0){$+$}}
\put(1018,289){\makebox(0,0){$+$}}
\put(1049,348){\makebox(0,0){$+$}}
\put(1080,392){\makebox(0,0){$+$}}
\put(1111,437){\makebox(0,0){$+$}}
\put(1142,476){\makebox(0,0){$+$}}
\put(1173,510){\makebox(0,0){$+$}}
\put(1204,543){\makebox(0,0){$+$}}
\put(1235,576){\makebox(0,0){$+$}}
\put(1266,595){\makebox(0,0){$+$}}
\put(1297,618){\makebox(0,0){$+$}}
\put(1328,642){\makebox(0,0){$+$}}
\put(1359,672){\makebox(0,0){$+$}}
\put(1390,687){\makebox(0,0){$+$}}
\put(1421,710){\makebox(0,0){$+$}}
\end{picture}
\end{center}
\caption{(a) Average lifetime of a packet $\overline{\protect\tau}(k)$ as a
function of $\protect\lambda$ for $k=1500$ and for $\mathcal{L}_l^{np}(50)$
lattice with $d_{M}$ metric and $m=D_{max}$, and with the number of extra
random links $l=0$ ($\circ$), $l=100$ ($\bullet$), $l=200$ ($\star$), and $%
l=400$ ($+$). (b) The same plot for $\mathcal{L}_l^{p}(50)$ lattice with $%
d_{PM}$ metric and $m=D_{max}$.}
\label{fig400}
\end{figure}
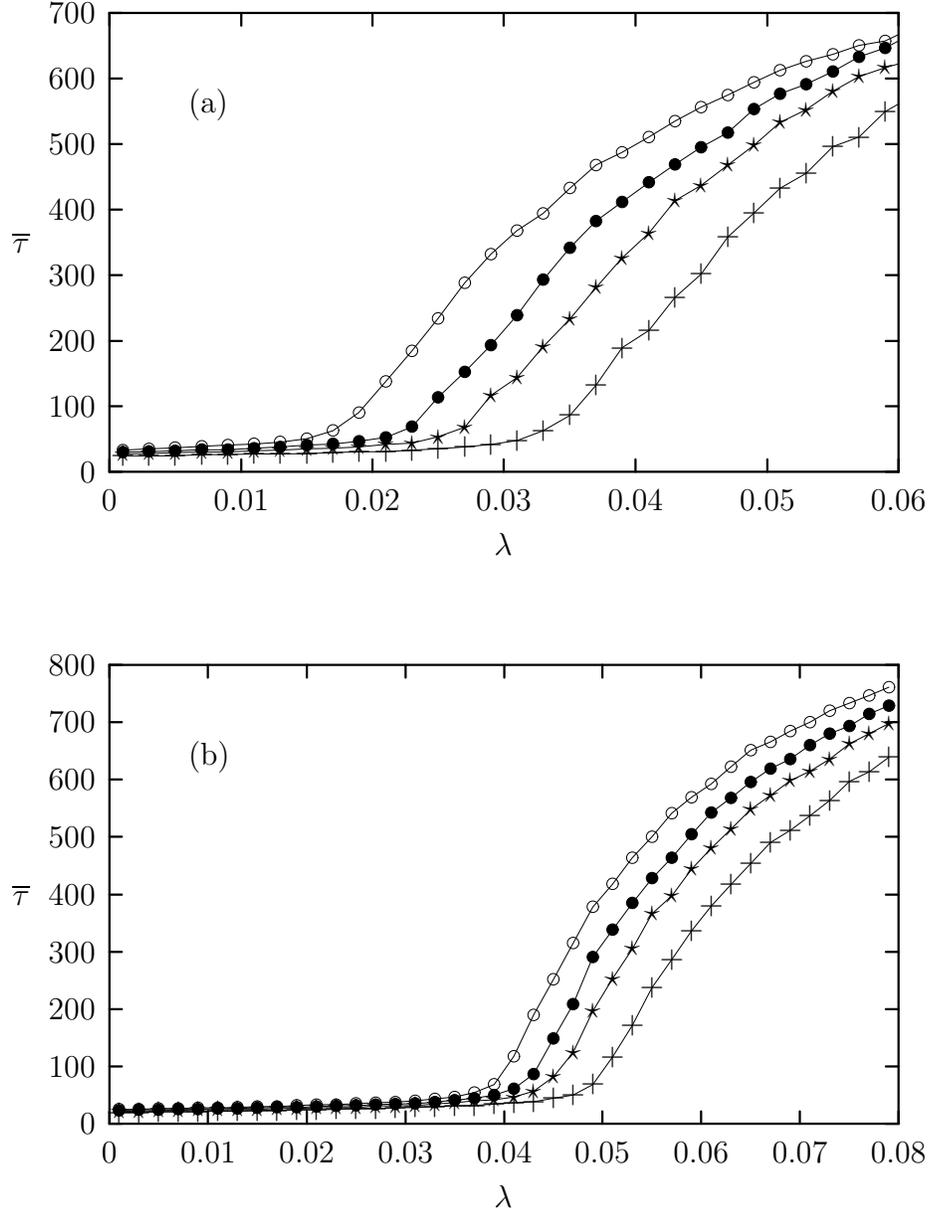

\begin{figure}[p]
\begin{center}
\setlength{\unitlength}{0.240900pt}
\begin{picture}(1500,900)(0,0)
\tenrm
\thicklines \path(241,134)(261,134)
\thicklines \path(1436,134)(1416,134)
\put(219,134){\makebox(0,0)[r]{0.01}}
\thicklines \path(241,206)(261,206)
\thicklines \path(1436,206)(1416,206)
\put(219,206){\makebox(0,0)[r]{0.015}}
\thicklines \path(241,278)(261,278)
\thicklines \path(1436,278)(1416,278)
\put(219,278){\makebox(0,0)[r]{0.02}}
\thicklines \path(241,350)(261,350)
\thicklines \path(1436,350)(1416,350)
\put(219,350){\makebox(0,0)[r]{0.025}}
\thicklines \path(241,422)(261,422)
\thicklines \path(1436,422)(1416,422)
\put(219,422){\makebox(0,0)[r]{0.03}}
\thicklines \path(241,495)(261,495)
\thicklines \path(1436,495)(1416,495)
\put(219,495){\makebox(0,0)[r]{0.035}}
\thicklines \path(241,567)(261,567)
\thicklines \path(1436,567)(1416,567)
\put(219,567){\makebox(0,0)[r]{0.04}}
\thicklines \path(241,639)(261,639)
\thicklines \path(1436,639)(1416,639)
\put(219,639){\makebox(0,0)[r]{0.045}}
\thicklines \path(241,711)(261,711)
\thicklines \path(1436,711)(1416,711)
\put(219,711){\makebox(0,0)[r]{0.05}}
\thicklines \path(241,783)(261,783)
\thicklines \path(1436,783)(1416,783)
\put(219,783){\makebox(0,0)[r]{0.055}}
\thicklines \path(241,855)(261,855)
\thicklines \path(1436,855)(1416,855)
\put(219,855){\makebox(0,0)[r]{0.06}}
\thicklines \path(241,134)(241,154)
\thicklines \path(241,855)(241,835)
\put(241,89){\makebox(0,0){1}}
\thicklines \path(361,134)(361,144)
\thicklines \path(361,855)(361,845)
\thicklines \path(431,134)(431,144)
\thicklines \path(431,855)(431,845)
\thicklines \path(481,134)(481,144)
\thicklines \path(481,855)(481,845)
\thicklines \path(519,134)(519,144)
\thicklines \path(519,855)(519,845)
\thicklines \path(551,134)(551,144)
\thicklines \path(551,855)(551,845)
\thicklines \path(578,134)(578,144)
\thicklines \path(578,855)(578,845)
\thicklines \path(601,134)(601,144)
\thicklines \path(601,855)(601,845)
\thicklines \path(621,134)(621,144)
\thicklines \path(621,855)(621,845)
\thicklines \path(639,134)(639,154)
\thicklines \path(639,855)(639,835)
\put(639,89){\makebox(0,0){10}}
\thicklines \path(759,134)(759,144)
\thicklines \path(759,855)(759,845)
\thicklines \path(829,134)(829,144)
\thicklines \path(829,855)(829,845)
\thicklines \path(879,134)(879,144)
\thicklines \path(879,855)(879,845)
\thicklines \path(918,134)(918,144)
\thicklines \path(918,855)(918,845)
\thicklines \path(949,134)(949,144)
\thicklines \path(949,855)(949,845)
\thicklines \path(976,134)(976,144)
\thicklines \path(976,855)(976,845)
\thicklines \path(999,134)(999,144)
\thicklines \path(999,855)(999,845)
\thicklines \path(1019,134)(1019,144)
\thicklines \path(1019,855)(1019,845)
\thicklines \path(1038,134)(1038,154)
\thicklines \path(1038,855)(1038,835)
\put(1038,89){\makebox(0,0){100}}
\thicklines \path(1158,134)(1158,144)
\thicklines \path(1158,855)(1158,845)
\thicklines \path(1228,134)(1228,144)
\thicklines \path(1228,855)(1228,845)
\thicklines \path(1277,134)(1277,144)
\thicklines \path(1277,855)(1277,845)
\thicklines \path(1316,134)(1316,144)
\thicklines \path(1316,855)(1316,845)
\thicklines \path(1348,134)(1348,144)
\thicklines \path(1348,855)(1348,845)
\thicklines \path(1374,134)(1374,144)
\thicklines \path(1374,855)(1374,845)
\thicklines \path(1397,134)(1397,144)
\thicklines \path(1397,855)(1397,845)
\thicklines \path(1418,134)(1418,144)
\thicklines \path(1418,855)(1418,845)
\thicklines \path(1436,134)(1436,154)
\thicklines \path(1436,855)(1436,835)
\put(1436,89){\makebox(0,0){1000}}
\thicklines \path(241,134)(1436,134)(1436,855)(241,855)(241,134)
\put(45,494){\makebox(0,0)[l]{\shortstack{${\lambda_c}$}}}
\put(838,21){\makebox(0,0){${l}$}}
\put(361,574){\circle{18}}
\put(481,574){\circle{18}}
\put(601,581){\circle{18}}
\put(721,581){\circle{18}}
\put(841,595){\circle{18}}
\put(960,610){\circle{18}}
\put(1080,639){\circle{18}}
\put(1200,711){\circle{18}}
\put(1320,812){\circle{18}}
\put(361,293){\circle*{18}}
\put(481,293){\circle*{18}}
\put(601,293){\circle*{18}}
\put(721,307){\circle*{18}}
\put(841,336){\circle*{18}}
\put(960,365){\circle*{18}}
\put(1080,422){\circle*{18}}
\put(1200,509){\circle*{18}}
\put(1320,639){\circle*{18}}
\end{picture}
\end{center}
\caption{Critical load $\protect\lambda_c$ as a function of a number of
extra links $l$ for the lattice $50 \times 50$ with periodic ($\circ$)
and non-periodic ($\bullet$) boundaries, using $d_{PM}$ and $d_M$
metric, respectively. In both cases, $m=D_{max}$.}
\label{fig550}
\end{figure}
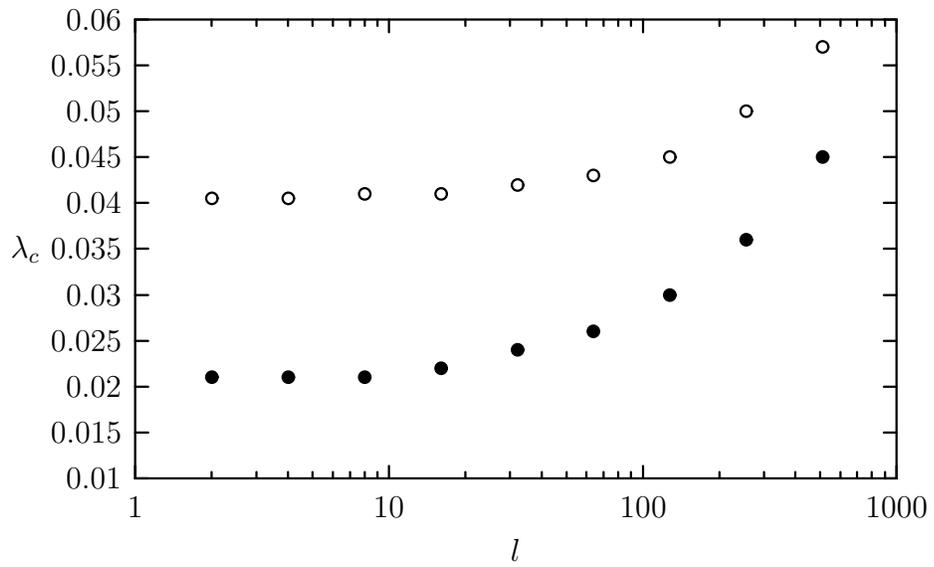


\begin{figure}[p]
\begin{center}
\vskip 0.9cm
\include{fig600}
\end{center}
\caption{Change in critical load for a lattice $\mathcal{L}^{p}_0(50)$ using
metrics $d_{PM}$ and $d_{SP}$ with $m=20$ and $m=50$. Vertical axis
corresponds to $\big(\protect\lambda_c(m,l)-\protect\lambda_c(m,0)\big)/%
\protect\lambda_c(m,0)$, where $\protect\lambda_c(m,l)$ is a critical load
at a given $m$ and $l$.}
\label{fig600}
\end{figure}
\begin{figure}[p]
\begin{center}
\hspace{0.5cm}$l=0$\hspace{3.5cm}$l=100$\\[0pt]
$k=0$\includegraphics[scale=0.3]{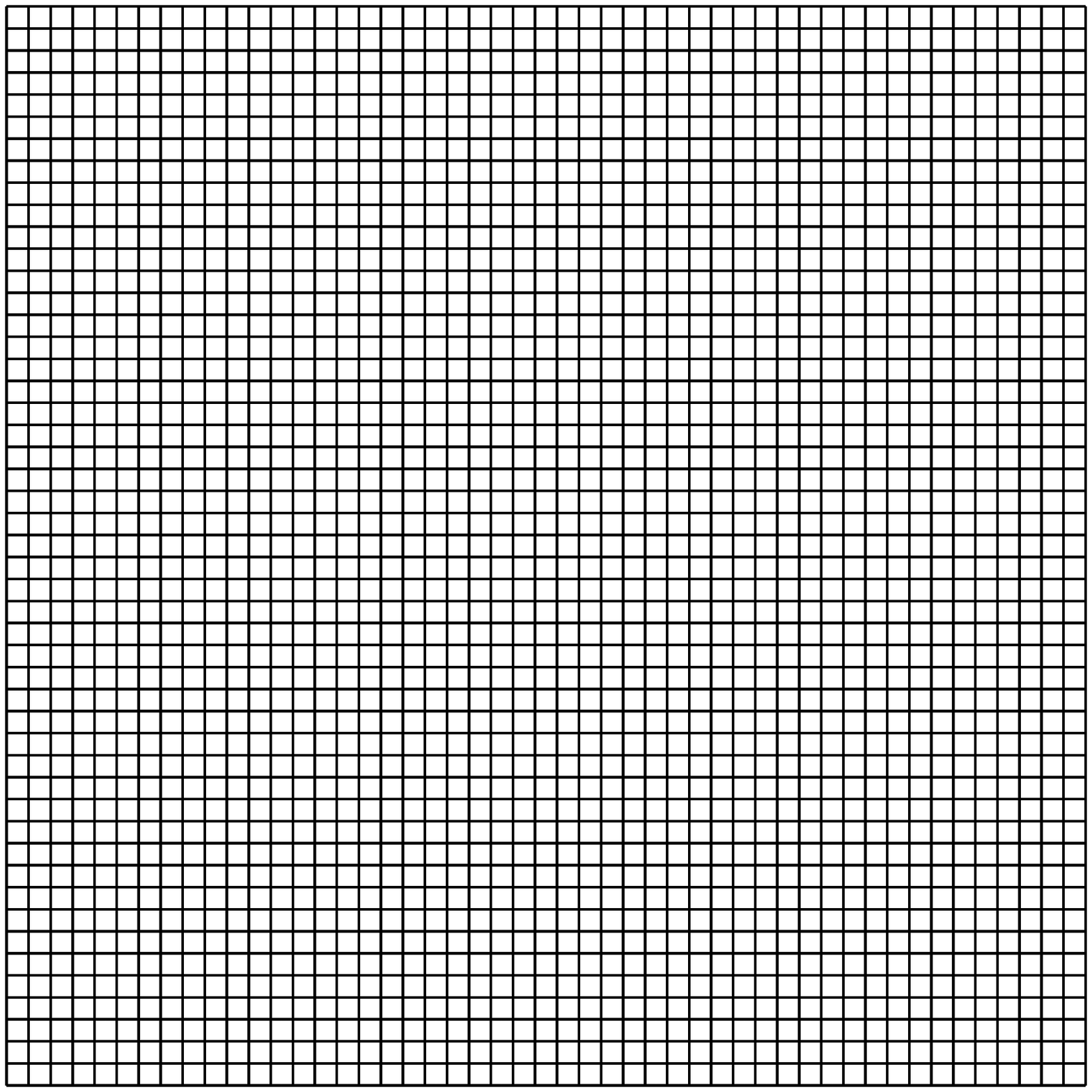} %
\includegraphics[scale=0.3]{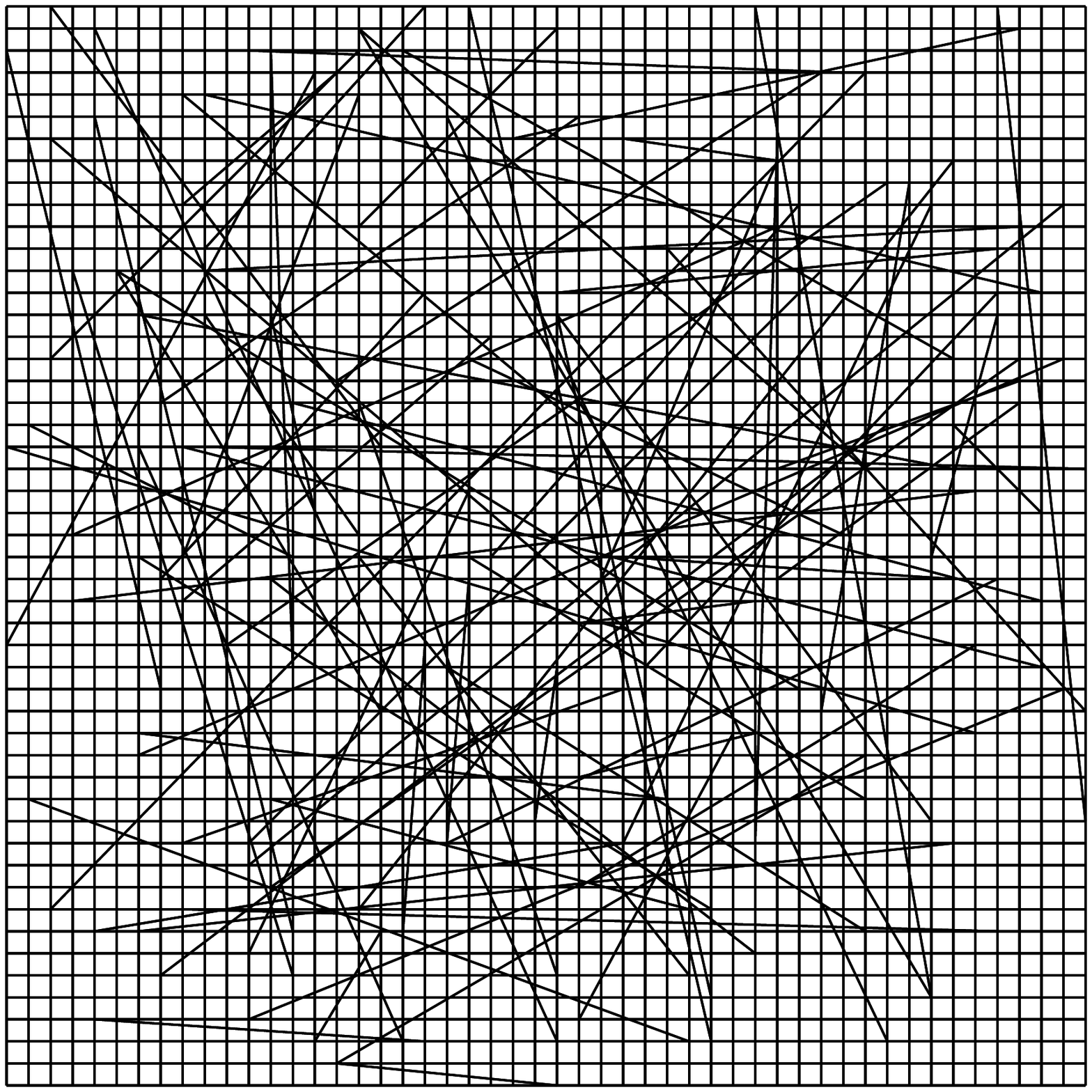}$k=0$\\[0pt]
$k=10^2$\includegraphics[scale=0.3]{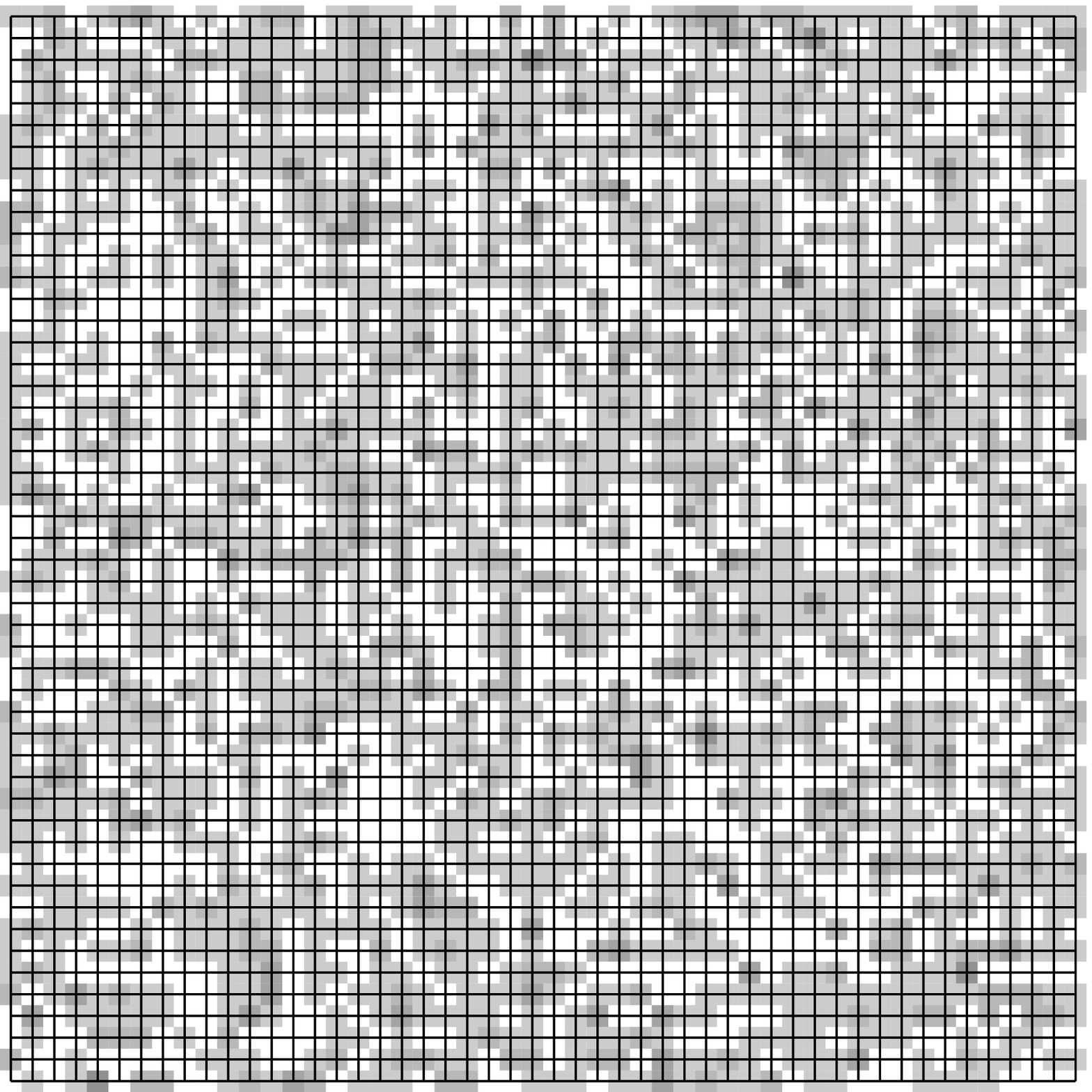} %
\includegraphics[scale=0.3]{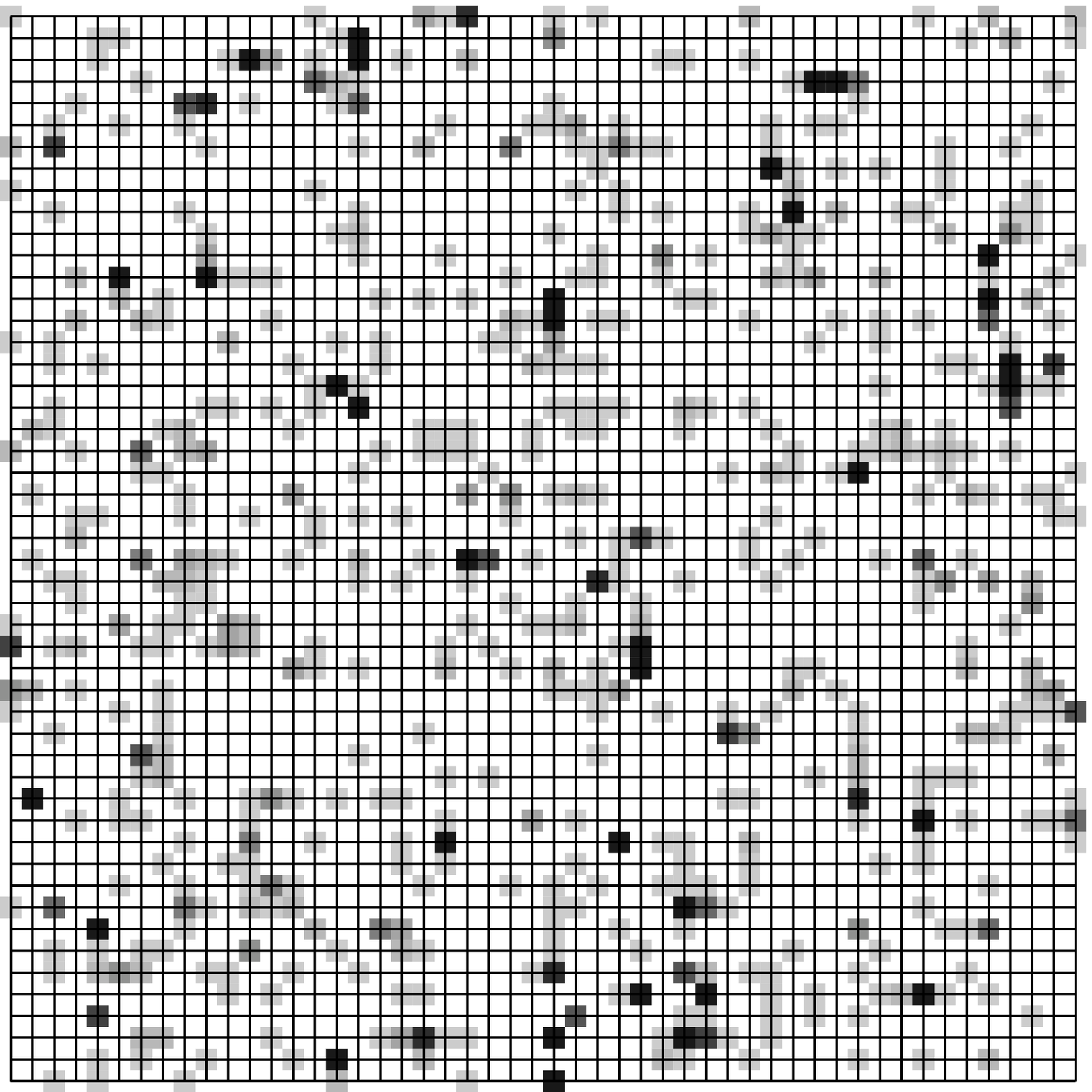}$k=10^2$\\[0pt]
$k=10^3$\includegraphics[scale=0.3]{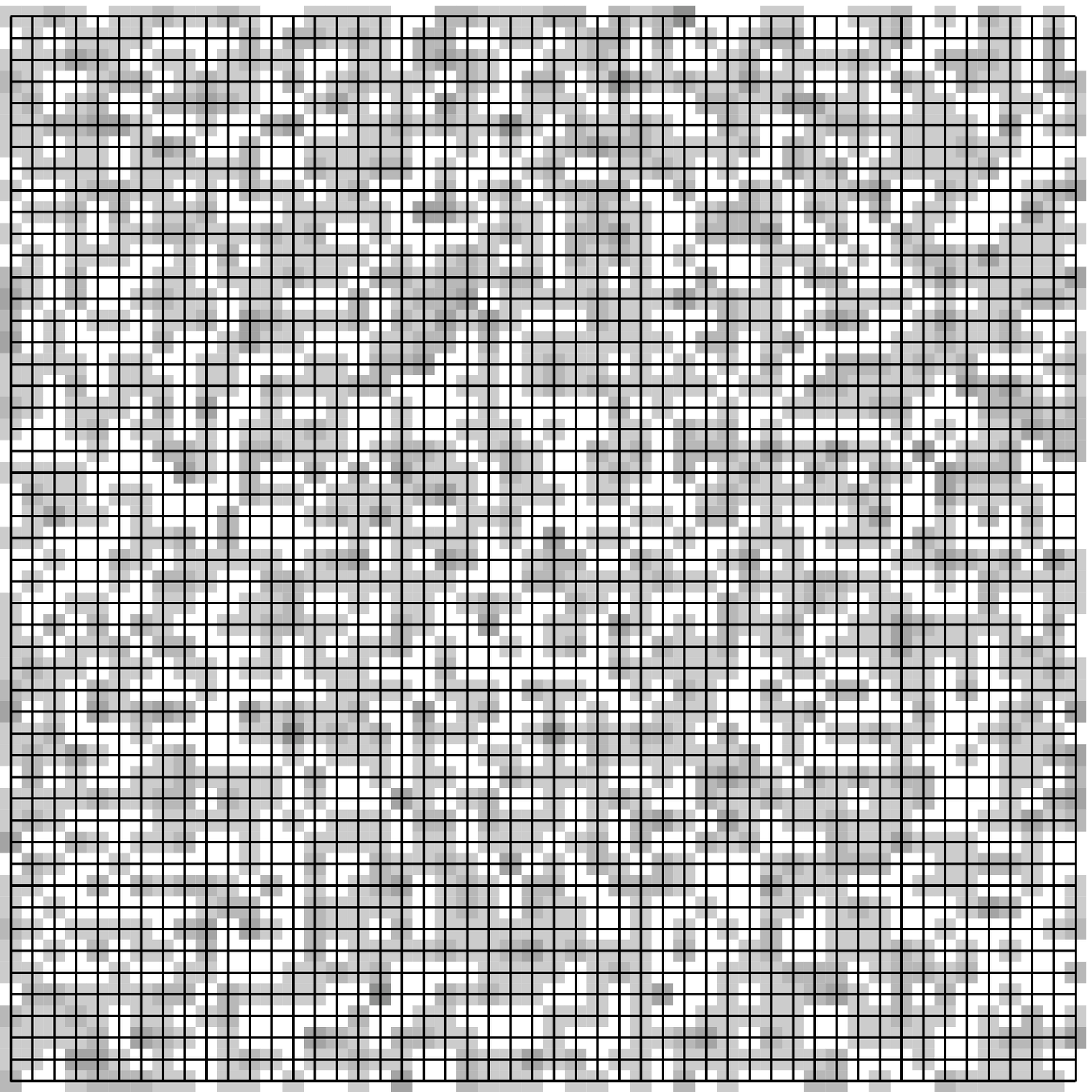} %
\includegraphics[scale=0.3]{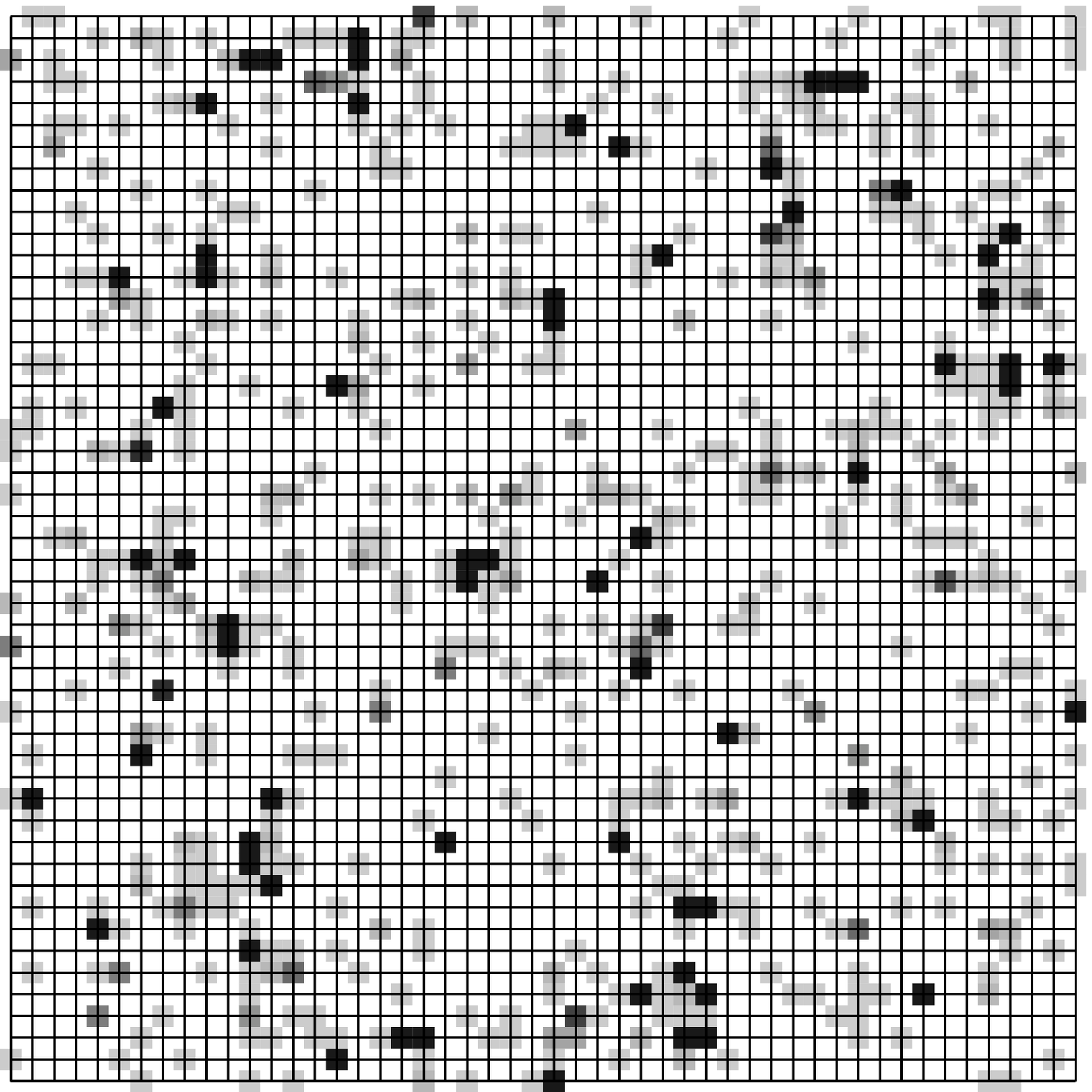}$k=10^3$%
\end{center}
\caption{Comparison of dynamics of the network $\mathcal{L}_l^p(50)$ with $%
d_{SP}$ and $\mathbf{\mathrm{R}}_\infty$ algorithm for $l=0$ (left column)
and $l=100$ (right column). Queue sizes are represented as shades of gray,
from the highest queue size of 20 or more represented by black color to the
empty queue represented by white color. In order to preserve clarity,
additional links are shown for $k=0$ only.}
\label{fig700}
\end{figure}
\begin{figure}[p]
\begin{center}
\hspace{0.5cm}$l=0$\hspace{3.5cm}$l=100$\\[0pt]
$k=0$\includegraphics[scale=0.3]{links0.ps} %
\includegraphics[scale=0.3]{links100.ps}$k=0$\\[0pt]
$k=10^2$\includegraphics[scale=0.3]{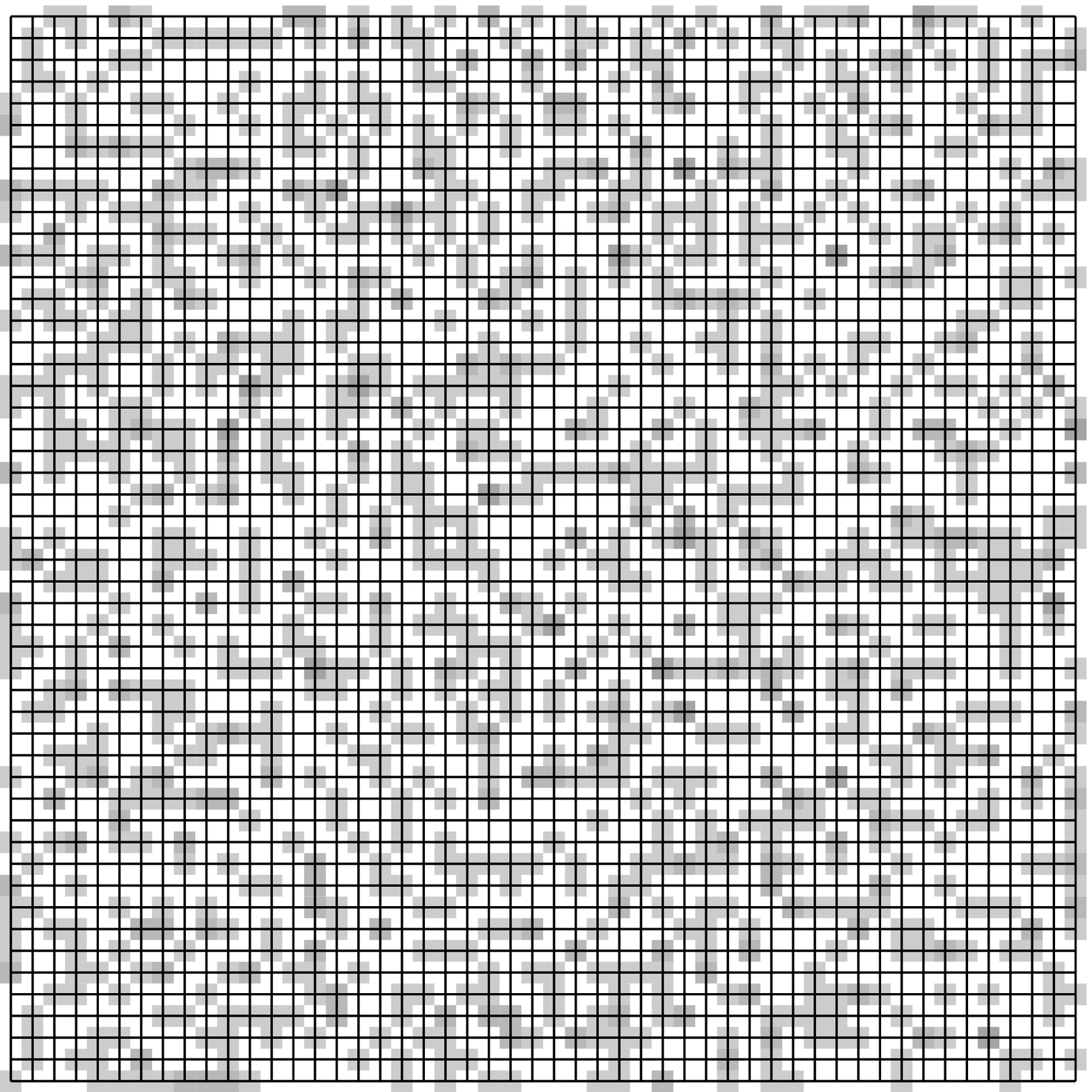} %
\includegraphics[scale=0.3]{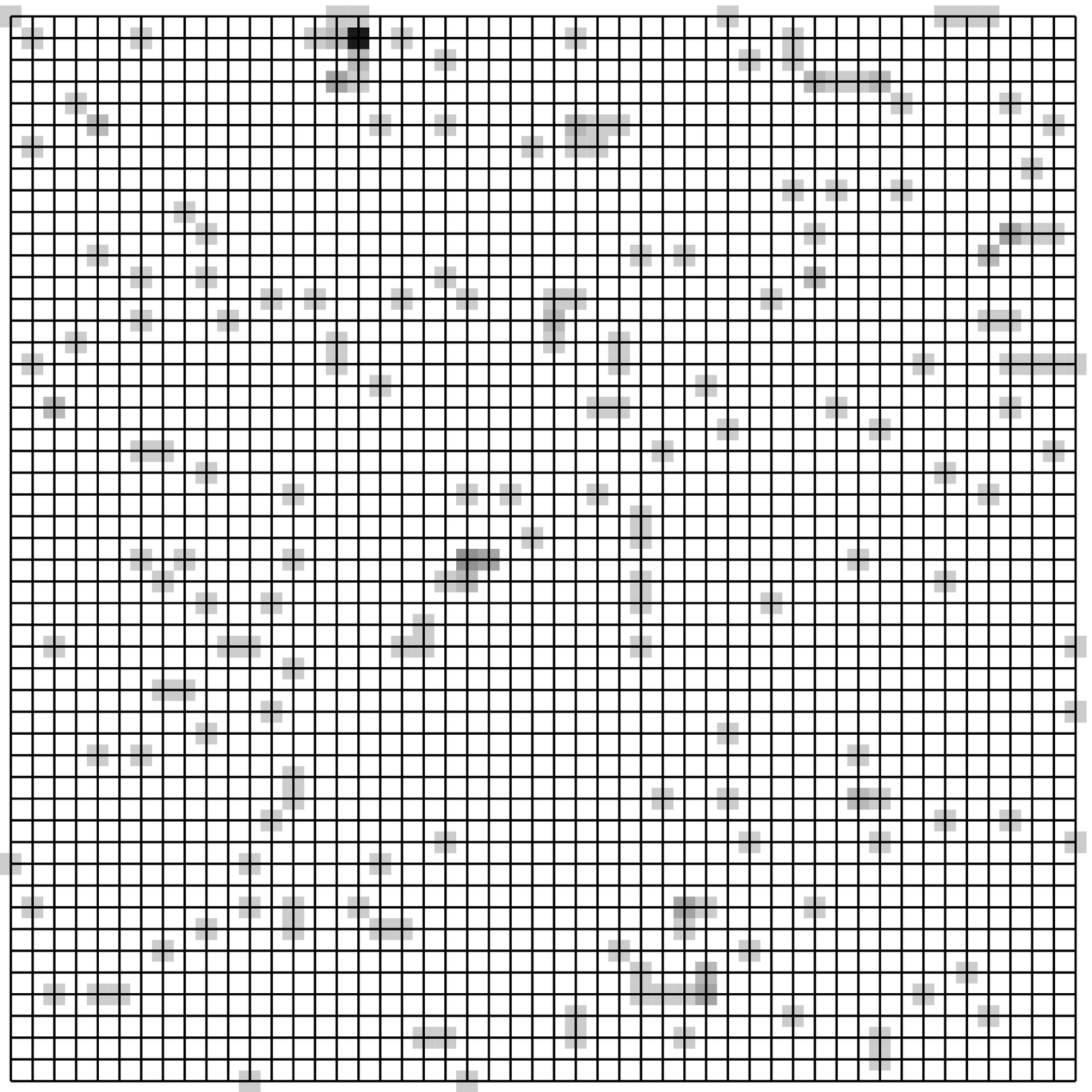}$k=10^2$\\[0pt]
$k=10^3$\includegraphics[scale=0.3]{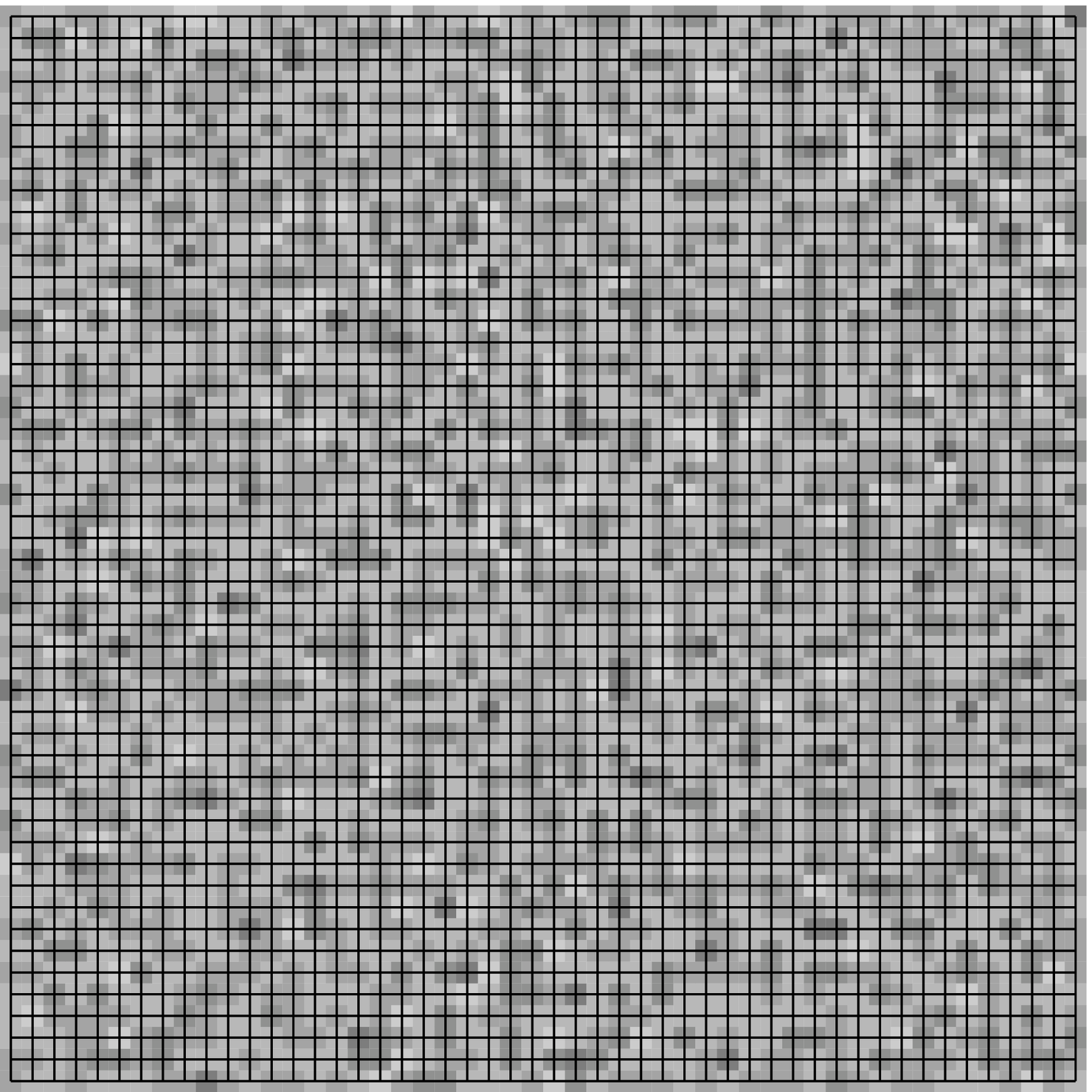} %
\includegraphics[scale=0.3]{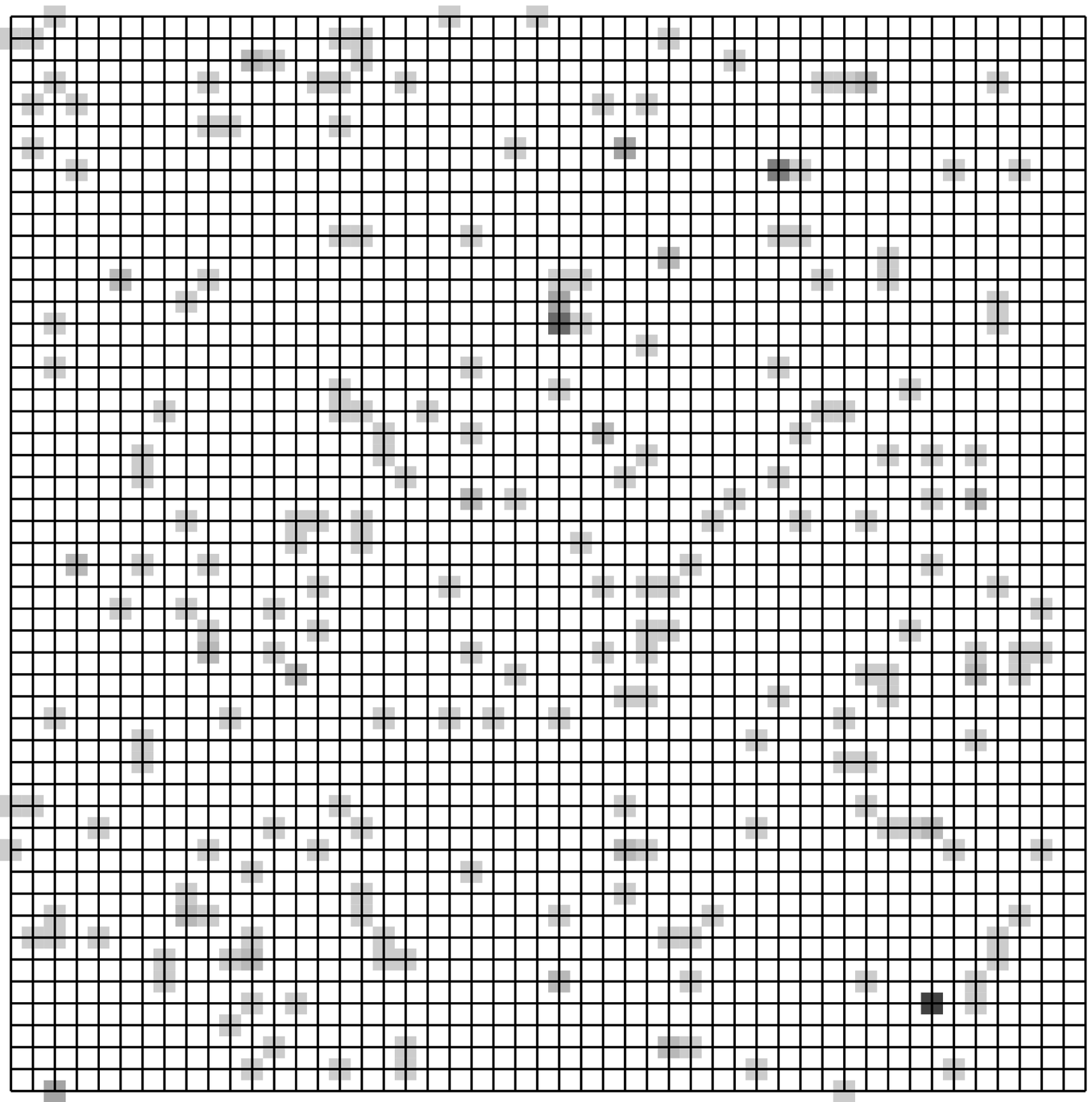}$k=10^3$%
\end{center}
\caption{Comparison of dynamics of the network $\mathcal{L}_l^p(50)$ with $%
d_{SP}$ and $\mathbf{\mathrm{R}}_{20}$ algorithm for $l=0$ (left column) and
$l=100$ (right column). Queue sizes are represented as shades of gray, from
the highest queue size of 20 or more represented by black color to the empty
queue represented by white color. In order to preserve clarity, additional
links are shown for $k=0$ only.}
\label{fig800}
\end{figure}

\end{document}